\documentclass[prd,aps,showpacs,nofootinbib,preprint]{revtex4}
\usepackage{graphicx}
\usepackage{amsfonts}
\usepackage{amssymb}
\usepackage{latexsym}

\newcommand{\CO}{{\cal O}}
\newcommand{\CL}{{\cal L}}
\newcommand{\CM}{{\cal M}}
\newcommand{\CN}{{\cal N}}
\newcommand{\CD}{{\cal D}}
\newcommand{\CE}{{\cal E}}
\newcommand{\bear}{\begin{array}}  \newcommand{\eear}{\end{array}}
\newcommand{\bea}{\begin{eqnarray}}  \newcommand{\eea}{\end{eqnarray}}
\newcommand{\beq}{\begin{equation}}  \newcommand{\eeq}{\end{equation}}
\newcommand{\bef}{\begin{figure}}  \newcommand{\eef}{\end{figure}}
\newcommand{\bec}{\begin{center}}  \newcommand{\eec}{\end{center}}
\newcommand{\non}{\nonumber}  
\newcommand{\lmk}{\left(}  \newcommand{\rmk}{\right)}
\newcommand{\lkk}{\left[}  \newcommand{\rkk}{\right]}
\newcommand{\lhk}{\left \{ }  \newcommand{\rhk}{\right \} }
\newcommand{\lnk}{\left \{ }  \newcommand{\rnk}{\right \} }
  
\newcommand{\del}{\partial}  
\newcommand{\vect}[1]{\mbox{\boldmath${#1}$}}
\newcommand{\vecs}[1]{\mbox{\boldmath\tiny${#1}$}}
\newcommand{\bib}{\bibitem} \newcommand{\new}{\newblock}
\newcommand{\la}{\left\langle} \newcommand{\ra}{\right\rangle}

\newcommand{\GeV}{{\rm GeV}}
\newcommand{\mg}{M_G}

\newcommand{\order}{\CO}

\newcommand{\Tr}{\mbox{Tr}}\newcommand{\tr}{\mbox{tr}}
\newcommand{\phip}{\phi_{+}}
\newcommand{\phim}{\phi_{-}}
\newcommand{\phic}{\phi_c}
\newcommand{\phid}{\phi_{\Delta}}

\newcommand{\im}{\mbox{Im}}
\newcommand{\re}{\mbox{Re}}
\newcommand{\step}{\theta(t_1-t_2)}
\newcommand{\phik}{\phi_{\vecs k}}
\newcommand{\Cak}{C_{a\vecs k}}
\newcommand{\Dak}{D_{a\vecs k}}
\newcommand{\xik}{\xi_{\vecs k}}
\newcommand{\vek}{\vect k}
\newcommand{\vep}{\vect p}
\newcommand{\vex}{\vect x}
\newcommand{\veg}{\vect \gamma}
\newcommand{\veq}{\vect q}
\newcommand{\veK}{\vect K}
\newcommand{\omk}{\omega_{k}}
\newcommand{\omp}{\omega_{p}}
\newcommand{\rhophi}{\rho_\phi}
\newcommand{\rhorad}{\rho_r}
\newcommand{\tildegs}{\tilde{g_\ast}}
\newcommand{\mgv}{m_{3/2}}
\newcommand{\mst}{M_\ast}

\def\PRD#1#2#3{Phys. Rev. D {\bf #1}, #2 (19#3)}

\begin{document}
\title{Fate of oscillating scalar fields in the thermal bath \\
and their cosmological implications}

\author{Jun'ichi Yokoyama} 
\affiliation{Department of Earth and Space Science, Graduate School of
  Science, Osaka University, Toyonaka 560-0043, Japan}

\date{\today}


\begin{abstract}
 Relaxation process of a coherent scalar field oscillation in the
 thermal bath is
 investigated using  nonequilibrium quantum field theory.
The Langevin-type equation of motion is obtained which has a memory term
 and both additive and multiplicative noise terms.
 The dissipation rate of the
 oscillating scalar field is calculated for various interactions such as
 Yukawa coupling, three-body scalar interaction, and biquadratic
 interaction.  When the background temperature is larger than the
 oscillation frequency, the dissipation rate arising from the
 interactions with
 fermions is suppressed due to the Pauli blocking, while it is enhanced
 for interactions with bosons due to the induced effect.
In both cases, we find that the microphysical detailed balance relation
 drives the oscillating field to a thermal equilibrium state.  That is, for
 low-momentum modes, the classical fluctuation-dissipation theorem holds
 and they relax to a state the equipartition law is satisfied, while
 higher-momentum modes reach the state the number density of each quanta
 consists of the thermal boson distribution function and zero-point
 vacuum contribution.
The temperature-dependent dissipation rates obtained here
are applied to the late
 reheating phase of inflationary universe.  It is found that in
 some cases the reheat
 temperature may take somewhat different value from the conventional
 estimates, and in an extreme case the inflaton can dissipate its energy
 without linear interactions that leads to its decay. 
Furthermore the evaporation rate of the Affleck-Dine field at the onset
 of its oscillation is calculated.

\end{abstract}

\pacs{98.80.Cq,11.10.Wx,05.40.-a \hspace{1.5cm} OU-TAP-232}

\maketitle

\tighten

\section{Introduction}

\label{sec:intro}

Cosmology of the early Universe is a useful probe of high energy 
phenomena beyond the reach of ground based
accelerator experiments.  The  universe at its birth,
however, is likely to suffer from huge relic quantum fluctuations and we
cannot expect that it started  classical evolution from a thermal
equilibrium state with a well-defined  temperature.  Rapid cosmic
expansion in the early universe further delays equilibration
\cite{Ellis:1979nq} and it is not likely that the phase transition of
grand unified theories occurred thermally \cite{Yokoyama:1989pa}.
Once the energy scale has fallen well below typical grand unification
scale, cosmic expansion rate  gets smaller than  interaction rates
of ambient massless particles to establish thermal equilibrium.
Phenomenon in such a regime may be studied in terms of quantum field
theory at finite temperature neglecting cosmic expansion and using the
cosmic temperature at each epoch.  If some degrees of freedom are out of
equilibrium, then we must of course use nonequilibrium field theories
\cite{Chou:es}.  
In modern cosmology, we often encounter a situation some
scalar fields are in
nonequilibrium configuration interacting with thermal background.

Indeed scalar fields play central roles to explain
virtually everything we observe---overall homogeneity and isotropy 
as well as the origin of small density perturbation are attributed 
to inflation driven by an {\it inflaton} scalar field, huge entropy
carried by the cosmic microwave (and neutrino) background radiation 
to the reheating process by the decay of the inflaton
\cite{Sato:1980yn,lindebook}. 
 Furthermore the
observed baryon asymmetry and dark matter
may also originate in scalar fields such as
squarks and/or sleptons through the Affleck-Dine mechanism
\cite{Affleck:1984fy}
and formation
of Q-balls \cite{Enqvist:2003gh}.

Thus it is of utmost importance to clarify the evolution of scalar
fields in cosmic medium.
In the present paper we study the fate of a coherent scalar field
oscillation 
interacting with fermions or bosons, which are thermally populated, using
the nonequilibrium quantum field theory.  Such a situation is realized
in the late stage of reheating after inflation as well as in 
 the evolution of flat directions in supersymmetric theories which may
 be associated with
Affleck-Dine baryogenesis.

We start with a brief review of a field theoretic method appropriate to
analyze time evolution of the expectation value of a scalar field.
The standard quantum field theory, which is appropriate for
evaluating the transition amplitude from an `in' state to an `out'
state for some field operator $\vect\CO$, $\la \rm out|\,\vect\CO\,|in
\ra$, is not suitable to trace time evolution of an expectation value
in a non-equilibrium system. In order to follow the time development
of the expectation value of some fields, it is necessary to establish
an appropriate extension of the quantum field theory, which is often
called the in-in formalism. This was first done by Schwinger
\cite{Sch} and developed in \cite{Bak,Kel,Jordan:ug}.
This method has been applied to various cosmological problems by a
number of authors \cite{Mor,Calzetta:1986ey,Boyanovsky:vi,
GR,Boyanovsky:1994me,
Yamaguchi:1996dp,GM,Yamaguchi:1997sy,Yokoyama:1998ju,Ramsey:1997sa}. 
 To name a few, a Langevin
equation has been obtained by Morikawa \cite{Mor} and Gleiser and Ramos
\cite{GR} in the slow-roll limit, which was applied to the electroweak
phase transition in \cite{Yamaguchi:1996dp} and to warm inflation \cite{Berera:1995ie}
 in \cite{Yokoyama:1998ju}.  On
the other hand, the case of oscillating scalar field was studied by
Greiner and M\"uller who took only the self interaction into account \cite{GM}.
Our work is partially related to it but we consider more general
interactions with other fermions and bosons, whose effects are
strikingly different from each other as shown in \cite{Yamaguchi:1996dp}.

We calculate an effective action for a real scalar field $\phi$
perturbatively in the
in-in formalism by integrating out  fields interacting with $\phi$
assuming that they are in thermal equilibrium distributions at a fixed
temperature in a fixed flat spacetime.  The resultant effective action
is complex-valued as a result of coarse graining of these interacting
fields, and it describes dissipation of the system field $\phi$. This
complex-valuedness is cured by the introduction of auxiliary fields
which act as noise terms, both additive and multiplicative, in the
equation of motion.  Its  derivation
 from the effective action is reported in the
next section.  

In \S III the equation of motion is explicitly solved in the case only
linear terms in $\phi$ are important.  We show that each spatial 
Fourier mode of the scalar field will relax to a value determined by the
ratio of the Fourier transform of the noise correlation function and
that of the memory kernel in the equation of motion, and it takes
 the same thermal equilibrium
value for all the three interactions discussed there, namely 
Yukawa coupling, three-body
scalar interaction, and biquadratic interaction.  This is achieved by
the detailed balance relation which also leads to the classical
fluctuation-dissipation theorem for  low momentum modes.
The time scale for the relaxation, which is essentially important for
cosmological applications, is also evaluated for respective
interactions.  The result is quite different depending on the
statistical property of the interacting particles.

In \S IV the analysis is extended to the multiplicative noises and
dissipation.  Although we cannot find a solution to the equation of
motion in this case, we can still confirm the generalized
fluctuation-dissipation relation and obtain the dissipation rate as
well.

These formulae are applied to two cosmological situations, namely the
late reheating phase after inflation in \S V and oscillating flat
direction in \S VI.  Finally \S VII is devoted to summary and discussion.

\section{Effective Action in Nonequilibrium Quantum Field Theory}
\subsection{Nonequilibrium quantum field theory}

We consider the following Lagrangian density of a singlet scalar
field $\phi$ interacting with another scalar field $\chi$ and a
fermion $\psi$.
\beq
   \CL = \frac12\,(\del_{\mu}\phi)^2-\frac12\,m_{\phi}^2\phi^2 -
     \frac{1}{4!}\lambda\,\phi^4
       + \frac12\,(\del_{\mu}\chi)^2-\frac12\,m_{\chi}^2\chi^2 -
    \CM \phi\chi^2 - 
     \frac{1}{4}\,h^2 \chi^2 \phi^2
       + i\bar\psi\gamma^{\mu}\del_{\mu}\psi - m_{\psi} \bar\psi \psi
     - f\phi\bar \psi\psi \:.  \label{lagrangian}
\eeq

When we investigate the time evolution of $\phi$, only the
initial condition is fixed, and so the time contour in a generating
functional starting from the infinite past must run to the infinite
future without fixing the final condition and come back to the
infinite past again. The generating functional in the in-in formalism
is thus given by
\bea
  Z[J,K,\eta,\bar\eta] &\equiv&
      \Tr \lkk T_p\lnk \exp
        \lkk i\int_{c}dt\int d^3x (J \phi+K \chi +\eta\psi+\bar\eta \bar\psi)
          \rkk \rnk \rho\,\rkk \non \\
       &=&
          \Tr \lkk T_{-}\lnk
            \exp\lkk i\int_{\infty}^{-\infty}dt \int d^3x
(J_{-}\phi_{-}+K_{-}\chi_{-}+\eta_{-}\psi_{-}
                +\bar\eta_{-}\bar\psi_{-})
                    \rkk \rnk \right. \non \\
       && ~~  \left.   \times\,
          T_{+}\lnk \exp\lkk i\int_{-\infty}^{\infty}dt \int d^3x
                (J_{+}\phi_{+}+K_{+}\chi_{+}+\eta_{+}\psi_{+}+
                    \bar\eta_{+} \bar\psi_{+})
               \rkk \rnk \rho\,\rkk \:, \non \\
\eea
where the suffix $c$ represents the closed time contour of
integration. $X_{+}$ denotes a field component $X$ on the plus-branch
($-\infty$ to $+\infty$) and $X_{-}$ that on the minus-branch ($+\infty$
to $-\infty$). The symbol $T_p$ represents the time ordering according
to the closed time contour, namely, $T_{+}$ the ordinary time ordering, and
$T_{-}$ the anti-time ordering. $J, K,$  and $\eta,~ \bar\eta$ represent
the external fields for the scalar and the Dirac fields,
respectively. In fact, each external field
$J_{+}(K_{+},\eta_{+},\bar\eta_{+})$ and
$J_{-}(K_{-},\eta_{-},\bar\eta_{-})$ is identical, 
but for technical
reasons we treat them differently and set
$J_{+}=J_{-}(K_{+}=K_{-},\eta_{+}=\eta_{-},\bar\eta_{+}=\bar\eta_{-})$
only at the end of calculation. $\rho$ is the initial density matrix.
Strictly speaking, we should couple the time development of the
expectation value of the field with that of the density matrix, which
is practically impossible. Accordingly we assume that deviation from
thermal equilibrium is small and use the density matrix corresponding to
the finite-temperature  state with the exception that the low-momentum modes
 of $\phi$ may have a larger amplitude initially, whose fate we
are interested in. Then the generating functional is
described by the path integral as 
\bea
  Z[\,J,K,\eta,\bar\eta\,]
    &=& \exp \biggl(\,iW[\,J,K,\eta,\bar\eta\,]\,\biggr) \non\\
     &=& \int_{c}\CD\phi \int_{c}\CD\chi \int_{c}\CD\psi
        \int_{c}\CD\psi^{\ast}
        \exp{\biggl(\,iS[\,\phi,\chi,\psi,\bar\psi,
J,K,\eta,\bar\eta\,]\,\biggr)} \:,
\eea
 where the classical action $S$ is given by \beq
  S[\,\phi,\chi,\psi,\bar\psi,J,K,\eta,\bar\eta\,]=
              \int_{c}d^4x \lkk \CL+J(x)\phi(x)+K(x)\chi(x)
                 +\eta(x)\psi(x)+\bar\eta(x)\bar\psi(x) \rkk \:.
\eeq 
As with the Euclidean-time formulation, the scalar field is
periodic and the Dirac field anti-periodic along the imaginary time
direction, with $\phi(t,\vect x)=\phi(t-i\beta,\vect x)$,
$\chi(t,\vect x)=\chi(t-i\beta,\vect x)$, and $\psi(t,\vect
x)=-\psi(t-i\beta,\vect x)$.  Here $\beta$ is the reciprocal of the
temperature $T$.

The effective action for the scalar field is defined by the connected
generating functional as 
\beq
\Gamma[\phi]=W[\,J,K,\eta,\bar\eta\,]-\int_{c}d^4x J(x)\phi(x) \:,
   \label{eqn:a1effe}
\eeq 
where $\phi(x)=\delta W[J,K,\eta,\bar\eta] / \delta J(x)$.
In terms of the components along the plus and the minus branches,
it reads
\beq
\Gamma[\phip,\phim]=W[J_+,J_-,\cdots]-\int_{-\infty}^{\infty}dt \int
d^3x \lkk J_+(x)\phip (x) -J_-(x)\phim (x)\rkk,
\eeq
with $\phip (x)=\delta W[J_+,J_-,\cdots] / \delta J_+(x)$ and
 $\phim (x)=-\delta W[J_+,J_-,\cdots] / \delta J_-(x)$.

We give the finite-temperature  propagator before the perturbative
expansion. For the closed path, the scalar propagator of $\chi$ has four
components consisting of $\chi_\pm (x)$ and $\chi_\pm (x')$.
\bea G_{\chi}(x-x') &=& \left(
    \begin{array}{cc}
       G^{++}_{\chi}(x-x') & G^{+-}_{\chi}(x-x')        \\
       G^{-+}_{\chi}(x-x') & G^{--}_{\chi}(x-x') \\
    \end{array}
   \right) \non \\
                    &\equiv&
   \left(
    \begin{array}{cc}
      \mbox{Tr}[\,T_{+}\chi_+(x)\chi_+(x')\rho\,] & 
      \mbox{Tr}[\,\chi_-(x')\chi_+(x)\rho\,] \\
      \mbox{Tr}[\,\chi_-(x)\chi_+(x')\rho\,] & 
      \mbox{Tr}[\,T_{-}\chi_-(x)\chi_-(x')\rho\,] \\
    \end{array}
   \right) \non \\
&=& \left(
    \begin{array}{cc}
       G^{F}_{\chi}(x-x') & G^{+}_{\chi}(x-x')        \\
       G^{-}_{\chi}(x-x') & G^{\tilde F}_{\chi}(x-x') \\
    \end{array}
   \right) \non \\
 &\equiv& 
   \int \frac{d^4 k}{(2\pi)^4}e^{-ik(x-x')}
   \left(
    \begin{array}{cc}
       G^{F}_{\chi}(k) & G^{+}_{\chi}(k)        \\
       G^{-}_{\chi}(k) & G^{\tilde F}_{\chi}(k) \\
    \end{array}
  \right) \:, 
\eea 
where
\bea
  G^{F}_{\chi}(k) &=& \frac{i}{k^2-m_{\chi}^2+i\epsilon}
                          +2\pi n_B(\omk)   
                \,\delta(k^2-m_{\chi}^2) \:, \non \\
  G^{\tilde F}_{\chi}(k) &=& \frac{-i}{k^2-m_{\chi}^2-i\epsilon}
                                 +2\pi n_B(\omk)   
                \,\delta(k^2-m_{\chi}^2) \:, \non \\
  G^{+}_{\chi}(k) &=& 2\pi\,[\,\theta(-k_{0})+n_B(\omk)\,]   
                \,\delta(k^2-m_{\chi}^2) \:, \non \\
  G^{-}_{\chi}(k) &=& 2\pi\,[\,\theta(k_{0})+n_B(\omk)\,]   
                \,\delta(k^2-m_{\chi}^2) \:,
\eea
with $n_B(\omk)=(e^{\beta\omk}-1)^{-1}$,\,
$\omk=\sqrt{\vect k^2+m_{\chi}^2}$,\, 
and $\epsilon(k_{0}) = \theta(k_{0})-\theta(-k_{0})$ \cite{pro}.
Similar formulae apply for $\phi$ field as well.

The propagator for a Dirac fermion is given by
\bea S_{\psi}(x-x') &=& \left(
    \begin{array}{cc}
       S^{++}_{\psi}(x-x') & S^{+-}_{\psi}(x-x')        \\
       S^{-+}_{\psi}(x-x') & S^{--}_{\psi}(x-x') \\
    \end{array}
   \right) \non \\
                    &\equiv&
   \left(
    \begin{array}{cc}
      \mbox{Tr}[\,T_{+}\psi_+(x)\bar\psi_+(x')\rho\,] & 
      \mbox{Tr}[\,-\bar\psi_-(x')\psi_+(x)\rho\,] \\
      \mbox{Tr}[\,\psi_-(x)\bar\psi_+(x')\rho\,] & 
      \mbox{Tr}[\,T_{-}\psi_-(x)\bar\psi_-(x')\rho\,] \\
    \end{array}
   \right) \non \\  &\equiv& 
\left(
    \begin{array}{cc}
       S^{F}_{\psi}(x-x') & S^{+}_{\psi}(x-x')        \\
       S^{-}_{\psi}(x-x') & S^{\tilde F}_{\psi}(x-x') \\
    \end{array}
   \right) \non \\
                    &\equiv&
   \int \frac{d^4 k}{(2\pi)^4}e^{-ik(x-x')}
   \left(
    \begin{array}{cc}
       S^{F}_{\psi}(k) & S^{+}_{\psi}(k)        \\
       S^{-}_{\psi}(k) & S^{\tilde F}_{\psi}(k) \\
    \end{array}
  \right) \:, 
\eea 
where
\bea
  S^{F}_{\psi}(k) &=& \frac{i}{\not{k}-m_{\psi}+i\epsilon}
                       -2\pi n_F(E_{k})   
                (\not{k}+m_{\psi})\,\delta(k^2-m_{\psi}^2) \:, \non \\
  S^{\tilde F}_{\psi}(k) &=&  \frac{-i}{\not{k}-m_{\psi}-i\epsilon}
                               -2\pi n_F(E_{k})   
                (\not{k}+m_{\psi})\,\delta(k^2-m_{\psi}^2) \:, \non \\
  S^{+}_{\psi}(k) &=& 2\pi\, 
             [\,\theta(-k_{0})-n_F(E_{k})\,]\,(\not{k}+m_{\psi})
                \,\delta(k^2-m_{\psi}^2) \:, \non \\
  S^{-}_{\psi}(k) &=& 2\pi\,
             [\,\theta(k_{0})-n_F(E_{k})\,]\,(\not{k}+m_{\psi}) 
                \,\delta(k^2-m_{\psi}^2) \:,
\eea
with $n_F(E_{k})=(e^{\beta E_{k}}+1)^{-1},\, 
E_{k}\equiv \sqrt{\vect k^2+m_{\psi}^2}$ \cite{pro}.

\subsection{Perturbative expansion of the finite-temperature effective action}

\begin{figure}[htb]
\begin{center}
\includegraphics[width=16cm]{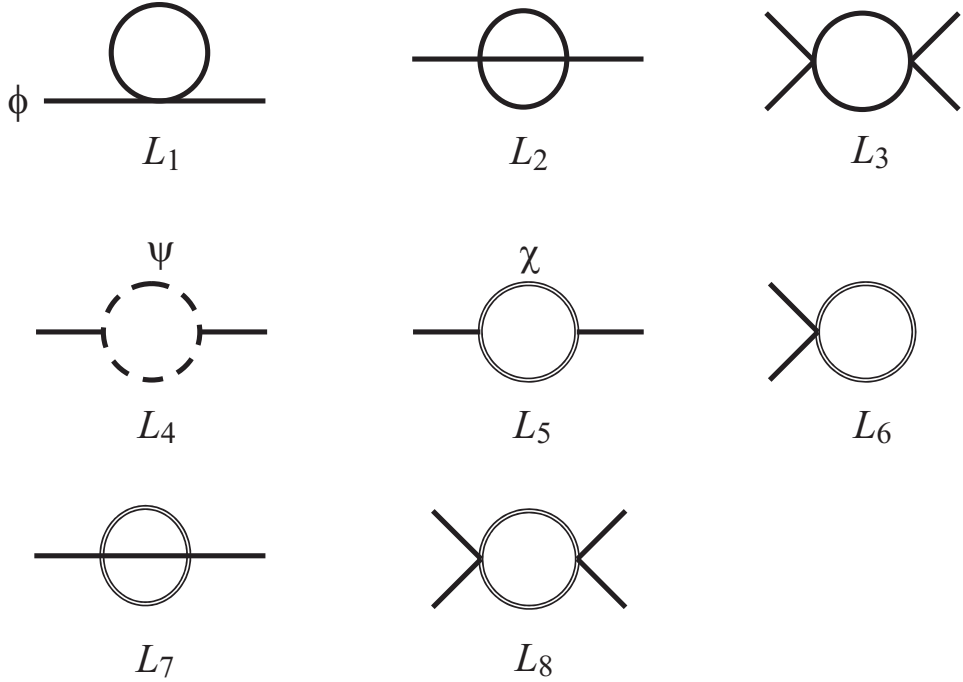}
\end{center}
\caption{Feynman diagrams corresponding to each term of the effective
 action.  Solid line denotes $\phi$, while broken line and double line
 represent $\psi$ and $\chi$, respectively.}
\label{fig:one}
\end{figure}

\indent
The perturbative loop expansion for the effective action $\Gamma$ can
be obtained by transforming $\phi \rightarrow \phi_{cl}+\zeta$
where $\phi_{cl}$ is the field configuration which extremizes the
classical action $S[\,\phi,J\,]$ and $\zeta$ is small perturbation
around $\phi_{cl}$. Up to two loop order and $\CO(\lambda^2, h^4,
f^2)$, $\Gamma$ is made up of the graphs as those depicted in Fig.\
\ref{fig:one} etc. Summing up these graphs, the effective action
$\Gamma$ becomes
\bea
 \Gamma[\phip,\phim]&=&\int d^4x\lkk \frac{1}{2}(\partial\phip)^2-
\frac{1}{2}(\partial\phim)^2-\frac{1}{2}m_\phi^2\lmk\phip^2(x)-\phim^2(x)\rmk
-\frac{\lambda}{4!}\lmk\phip^4(x)-\phim^4(x)\rmk\rkk \nonumber \\
&& +\sum_{j=1}^8  L_j[\phip,\phim]+\cdots, 
\eea
where each  of $ L_j[\phip,\phim]$ corresponds to each graph in
Fig.\ \ref{fig:one} and is given as follows.
\bea
 L_1&=&-\frac{\lambda}{4}\int d^4x G_\phi^{++}(0)
\lkk \phip^2(x)-\phim^2(x)\rkk,\\
 L_2&=&\frac{i\lambda^2}{12}\int d^4x d^4x'
\lkk \phip(x)G_\phi^{++}(x-x')^3\phip(x')
- \phim(x)G_\phi^{-+}(x-x')^3\phip(x')\right.  \nonumber\\
&&\qquad \left.- \phip(x)G_\phi^{+-}(x-x')^3\phim(x')
+ \phim(x)G_\phi^{--}(x-x')^3\phim(x')\rkk,\\
 L_3&=&\frac{i\lambda^2}{16}\int d^4x d^4x'
\lkk \phip^2(x)G_\phi^{++}(x-x')^2\phip^2(x')
-  \phim^2(x)G_\phi^{-+}(x-x')^2\phip^2(x')\right.  \nonumber\\
&&\qquad \left. -\phip^2(x)G_\phi^{+-}(x-x')^2\phim^2(x')
+\phim^2(x)G_\phi^{--}(x-x')^2\phim^2(x')\rkk,\\
 L_4&=&-\frac{if^2}{2}\int d^4x d^4x'
\lkk \phip(x)S^{++}_{\psi}(x-x')S^{++}_{\psi}(x'-x)\phip(x')
\right.  \nonumber\\
&&\qquad \left.
- \phim(x)S^{-+}_{\psi}(x-x')S^{+-}_{\psi}(x'-x)\phip(x')
-\phip(x)S^{+-}_{\psi}(x-x')S^{-+}_{\psi}(x'-x)\phim(x')
\right.  \nonumber\\
&&\qquad \left.
+\phim(x)S^{--}_{\psi}(x-x')S^{--}_{\psi}(x'-x)\phim(x')\rkk,\\
 L_5&=&i\CM^2\int d^4x d^4x' \lkk
\phip(x)G_\chi^{++}(x-x')^2\phip(x')
- \phim(x)G_\chi^{-+}(x-x')^2\phip(x')\right.  \nonumber\\
&&\qquad \left. - \phip(x)G_\chi^{+-}(x-x')^2\phim(x')
+ \phim(x)G_\chi^{--}(x-x')^2\phim(x')\rkk,\\
 L_6&=&-\frac{h^2}{4}\int d^4x G_\chi^{++}(0)
\lkk \phip^2(x)-\phim^2(x)\rkk,\\
 L_7&=&\frac{ih^4}{4}\int d^4x d^4x'
\lkk \phip(x)G_\phi^{++}(x-x')G_\chi^{++}(x-x')^2\phip(x')
\right.\non\\
&&\qquad -
\phim(x)G_\phi^{-+}(x-x')G_\chi^{-+}(x-x')^2\phip(x')
  \nonumber\\
&&\qquad - \phip(x)G_\phi^{+-}(x-x')
G_\chi^{+-}(x-x')^2\phim(x')  \nonumber\\
&&\qquad\left.
+ \phim(x)G_\phi^{--}(x-x')G_\chi^{--}(x-x')^2\phim(x')\rkk,\\
 L_8&=&\frac{ih^4}{16}\int d^4x d^4x'
\lkk \phip^2(x)G_\chi^{++}(x-x')^2\phip^2(x')
-  \phim^2(x)G_\chi^{-+}(x-x')^2\phip^2(x')\right.  \nonumber\\
&&\qquad \left. -\phip^2(x)G_\chi^{+-}(x-x')^2\phim^2(x')
+\phim^2(x)G_\chi^{--}(x-x')^2\phim^2(x')\rkk.
\eea
It is convenient to introduce new variables
\beq
 \phic\equiv \frac{1}{2}(\phip+\phim)~~{\rm and}~~\phid\equiv\phip-\phim
\eeq
to rewrite the effective action in terms of these variables.
As will be seen in (\ref{eqn:a3cla}), $\phi_\Delta$ is a response field
 and $\phi_c$ is the physical field. 
We find
\bea
 \Gamma[\phic,\phid]&=&\int d^4x\lkk -\phid(x)\Box\phic(x)
-m_\phi^2\phid(x)\phic(x)-\frac{\lambda}{4!}\lmk \phid^3(x)\phic(x)
+4\phid(x)\phic^3(x)\rmk
\rkk \nonumber \\
&& +\sum_{j=1}^8  L_j[\phic,\phid]+\cdots, 
\eea
with
\bea
 L_1&=&-\frac{\lambda}{2}\int d^4x
G_\phi^{++}(0)\phid(x)\phic(x),\\
 L_2&=&-\frac{\lambda^2}{3}\int d^4x d^4x'
\im\lkk  G_\phi^{++}(x-x')^3\rkk\phid(x)\phic(x')\step   \nonumber\\
&&
+\frac{i\lambda^2}{12}\int d^4x d^4x' \re\lkk 
 G_\phi^{++}(x-x')^3\rkk\phid(x)\phid(x'),  \label{22}\\
 L_3&=&-\frac{\lambda^2}{2}\int d^4x d^4x'
\im\lkk  G_\phi^{++}(x-x')^2\rkk\lmk
\phid(x)\phic(x)\phic^2(x')+
\textstyle{\frac{1}{4}}\phid(x)\phid^2(x')\phic(x)\rmk\step
  \nonumber\\
&& +\frac{i\lambda^2}{4}\int d^4x d^4x'
\re\lkk G_\phi^{++}(x-x')^2\rkk\phid(x)\phid(x')
\phic(x)\phic(x'),\\
 L_4&=& 2f^2\int d^4x d^4x'\im\lkk
S^{++}_{\psi}(x-x')S^{++}_{\psi}(x'-x)\rkk
\phid(x)\phic(x')\step \nonumber\\
&&
-\frac{if^2}{2}\int d^4x d^4x' \re\lkk
S^{++}_{\psi}(x-x')S^{++}_{\psi}(x'-x)\rkk
\phid(x)\phid(x'),\\
 L_5&=&-4\CM^2\int d^4x d^4x' \im\lkk
G_\chi^{++}(x-x')^2\rkk\phid(x)\phic(x')\step  \nonumber\\
&& +i\CM\int d^4x d^4x' \re\lkk G_\chi^{++}(x-x')^2\rkk
\phid(x)\phid(x'),\\
 L_6&=&-\frac{h^2}{2}\int d^4x
G_\chi^{++}(0)\phid(x)\phic(x),\\
 L_7&=&-h^4\int d^4x d^4x'
\im\lkk  G_\phi^{++}(x-x')G_\chi^{++}(x-x')^2\rkk
\phid(x)\phic(x')\step   \nonumber\\
&&
+\frac{ih^4}{4}\int d^4x d^4x' \re\lkk 
 G_\phi^{++}(x-x') G_\chi^{++}(x-x')^2\rkk\phid(x)\phid(x'), \\
 L_8&=&-\frac{h^4}{2}\int d^4x d^4x'
\im\lkk  G_\chi^{++}(x-x')^2\rkk\lkk
\phid(x)\phic(x)\phic^2(x')
+\textstyle{\frac{1}{4}}\phid(x)\phid^2(x')\phic(x)\rkk\step
  \nonumber\\
&& +\frac{ih^4}{4}\int d^4x d^4x'
\re\lkk G_\chi^{++}(x-x')^2\rkk\phid(x)\phid(x')
\phic(x)\phic(x'). \label{27}
\eea
Among these terms, $ L_1$ and $ L_6$ are corrections to the mass
term of $\phi$, while other terms have both real and imaginary parts.
As a result we find
\bea
&&  \Gamma[\phi_{c},\phi_{\Delta}] =\int d^4x \lnk
        -\phi_{\Delta}(x)[\,\Box+M^2\,]\phi_{c}(x)
        -\frac{\lambda}{4!}
           \lkk 
             4\phi_{\Delta}(x)\phi_{c}^3(x)+\phi_{c}(x)\phi_{\Delta}^3(x)  
           \rkk          
                                           \rnk \non \\
       && -\int d^4x d^4x' 
            \lkk\,  A_{2}(x-x')+A_{4}(x-x')+A_{5}(x-x')
	    +A_{7}(x-x')\, \rkk
 \theta(t_1-t_2)\phi_{\Delta}(x)\phi_{c}(x') \non \\
          && -\int d^4x d^4x' 
           \lkk\, A_{3}(x-x') + A_{8}(x-x') \,\rkk\theta(t_1-t_2)
	  \non \\ && \qquad\qquad\qquad 
\times\lkk
\phid(x)\phic(x)\phic^2(x')
+\textstyle{\frac{1}{4}}\phid(x)\phid^2(x')\phic(x)\rkk
            \non \\
       && +\frac{i}{2}\int d^4x d^4x' 
          \lhk\,
               \lkk B_{2}(x-x')+B_4(x-x')+B_5(x-x')+B_7(x-x')\rkk
\phi_{\Delta}(x)\phi_{\Delta}(x')\right. 
\non \\
 &&\left. \qquad \qquad \qquad 
+\lkk\, B_{3}(x-x') + B_{8}(x-x') \,\rkk
            \phi_{\Delta}(x)\phi_{\Delta}(x')\phi_{c}(x)\phi_{c}(x')  
\,\,\rhk  \:, 
  \label{eqn:a3effe}
\eea
where
\bea
    M^2 = m_\phi^2 +\lambda N(m_\phi) + 2h^2 N(m_\chi),~~~
  N(m_i)\equiv \int\!\frac{d^3q}{(2\pi)^3}
             \frac{1+2n_{B}(\omega_q)}
               {4\omega_q},~\omega_q\equiv
	     \sqrt{{\vect q}^2+m_i^2},
\eea 
\bea
A_{2}(x-x')  &=& \frac{\lambda^2}{3}
\im\lkk  G_\phi^{F}(x-x')^3\rkk \:, \label{32} \\
A_{3}(x-x')  &=& \frac{\lambda^2}{2}\im\lkk
G_\phi^{F}(x-x')^2\rkk \:, \label{33}\\
A_{4}(x-x')  &=& -2f^2\mbox{Im}\lnk\tr\lkk S^{F}_{\psi}(x-x')
S^{F}_{\psi}(x'-x)\rkk\rnk  \:,\label{34} \\
A_{5}(x-x')  &=& 4\CM^2 \im\lkk
G_\chi^{F}(x-x')^2\rkk  \:,\label{35} \\
A_{7}(x-x')  &=& h^4
\im\lkk  G_\phi^{F}(x-x')G_\chi^{F}(x-x')^2\rkk \:,\label{36} \\
A_{8}(x-x')  &=& \frac{h^4}{2}
\im\lkk  G_\chi^{F}(x-x')^2\rkk  \:,\label{37} \\
B_{2}(x-x')  &=& \frac{\lambda^2}{6} \re\lkk 
 G_\phi^{F}(x-x')^3\rkk \:,\label{38} \\
B_{3}(x-x')  &=& \frac{\lambda^2}{2}
\re\lkk G_\phi^{F}(x-x')^2\rkk \:,\label{39} \\
B_{4}(x-x')  &=& -f^2 \re\lnk\tr\lkk
S^{F}_{\psi}(x-x')S^{F}_{\psi}(x'-x)\rkk\rnk \:,\label{40} \\
B_{5}(x-x')  &=& 2\CM^2 \re\lkk G_\chi^{F}(x-x')^2\rkk \:,
\label{41}\\
B_{7}(x-x')  &=& \frac{h^4}{2}
\re\lkk  G_\phi^{F}(x-x')G_\chi^{F}(x-x')^2\rkk \:,\label{42} \\
B_{8}(x-x')  &=& \frac{h^4}{2}
\re\lkk G_\chi^{F}(x-x')^2\rkk \:. \label{43}
\eea
Apparently, $A_j$ and $B_j$ are related with the real and the imaginary 
parts of $ L_j$, respectively.  The above expressions for
$A_j(x-x')$ are valid only for $t_1-t_2 >0$.   We find
\bea
A_j(\vex'-\vex,t_1-t_2)&=&A_j(\vex-\vex',t_1-t_2),\label{parityeven}\\
A_j(\vex-\vex',t_2-t_1)&=&-A_j(\vex-\vex',t_1-t_2), \label{timeodd}
\eea
for $t_2-t_1 <0$, although only those with $t_1-t_2 >0$ appear
in the final expressions.  We also find
\bea
B_j(\vex'-\vex,t_1-t_2)&=&B_j(\vex-\vex',t_1-t_2), \label{parityevenb} \\
B_j(\vex-\vex',t_2-t_1)&=&B_j(\vex-\vex',t_1-t_2).\label{timeeven}
\eea

The imaginary parts of the effective action represent dissipative
effects and we can obtain real effective action by introducing 
auxiliary random
Gaussian fields, $\xi_a(x)$ and $\xi_m(x)$, as follows.
\beq
  \exp (i\Gamma[\phi_{c},\phi_{\Delta}])=\int\CD\xi_{a} \int\CD\xi_{m}
              P_{a}[\xi_{a}]P_{m}[\xi_{m}]\exp\lhk
           i\Gamma_{\rm eff}[\,\phi_{c},\phi_{\Delta},\xi_{a},\xi_{m}\,]\rhk
                    \:,
\eeq
where
\beq
  \Gamma_{\rm eff}[\,\phi_{c},\phi_{\Delta},\xi_{a},\xi_{m}\,] \equiv
                 \mbox{Re}\Gamma[\phi_{c},\phi_{\Delta}]
                  +\int d^4x[\,\xi_{a}(x)\phi_{\Delta}(x)
                    +\xi_{m}(x)\phi_{c}(x)\phi_{\Delta}(x)\,] \:.
 \label{eqn:a3cla}
\eeq
Here $P_{a}[\xi_{a}]$ and $P_{m}[\xi_{m}]$ are 
 the probability distribution functionals defined by
\bea
  P_{a}[\xi_{a}]~~~ &\equiv& \CN_{a} \exp \lkk\, - \frac12\int d^4x d^4x'
                      \xi_{a}(x)  D_{a}^{-1}(x-x')\xi_{a}(x')\,\rkk ,
\\
   D_a(x-x') &\equiv&  B_{2}(x-x')+B_4(x-x')+B_5(x-x')
  +B_7(x-x')     \:.
\\
  P_{m}[\xi_{m}]~~~ &\equiv& \CN_{m} \exp \lkk\, - \frac12\int d^4x d^4x'
                      \xi_{m}(x)  D_{m}^{-1}(x-x')\xi_{m}(x')\,\rkk ,
\\
   D_m(x-x') &\equiv& B_{3}(x-x')+B_8(x-x') ,    \:.
\eea
respectively.  Thus the dispersions of $\xi_a(x) $ and $\xi_m(x)$ are
given by $B_j$.  In the above expressions
 $\CN_{a}$ and $\CN_m$ are  normalization factors, while
 the inverse $D^{-1}_i(x-y)$ is defined by the relation
\beq
  \int d^4y D^{-1}_i(x-y)D_i(y-z)=\delta(x-z).
\eeq

\subsection{Equation of motion}
\label{subsec:lan}

\indent
Applying the variational principle to $\Gamma_{\rm eff}$, we obtain the
equation of motion for $\phi_{c}$ containing no imaginary quantity.
\beq
  \left.\frac{\delta \Gamma_{\rm eff}[\,\phi_{c},
\phi_{\Delta},\xi_{a},\xi_{m}\,]}
        {\delta \phi_{\Delta}}  
   \biggl. \right|_{\phi_{\Delta}=0}       
      =0 \:.
\eeq

\noindent
From (\ref{eqn:a3cla}), it reads
\bea
 && (\,\Box+M^2\,)\,\phi_{c}(x)
          +\frac{\lambda}{3!}\phi_{c}^3(x) 
      + \int_{-\infty}^{t}dt'\int d^3x'C_{a}(x-x')\phi_{c}(x') \non \\ 
&&\qquad\qquad\qquad+\phi_{c}(x)\int_{-\infty}^{t}dt'\int d^3x'
              C_{m}(x-x')\phi_{c}^2(x') 
           =\xi_{a}(x)+\phi_{c}(x)\xi_{m}(x) \:,
  \label{eqn:a4eqm}
\eea
with
\bea
 C_{a}(x-x')&\equiv&  A_{2}(x-x')+A_4(x-x')+A_5(x-x')+A_7(x-x'),  \\    
 C_{m}(x-x')&\equiv&  A_{3}(x-x')+A_8(x-x').
\eea
We shall call these two functions memory kernels because
the last two terms in the left-hand-side of (\ref{eqn:a4eqm})
are nonlocal in time.
They will reduce to the dissipation
terms and perturbative corrections to the classical equation of motion
which  would become  a part of the  derivative of the effective potential,
$V'_{\rm eff}(\phi)$, if we restricted $\phi(x')$ to be a constant in
space and time.   
In this equation of motion $\xi_a(x)$ should be regarded as
 an additive random Gaussian noise with the dispersion,
\beq
  \la\,\xi_{a}(x)\xi_{a}(x')\,\ra = D_{a}(x-x') \:,
  \label{eqn:corre}
\eeq
and $\xi_m(x)$ is a multiplicative random Gaussian
noise acting on $\phic(x)$ with the
dispersion, 
\beq
  \la\,\xi_{m}(x)\xi_{m}(x')\,\ra = D_{m}(x-x') \:.
  \label{eqn:corrm}
\eeq

\section{Analysis in the linear regime} \label{linearsection}

\subsection{Equation of motion in the Fourier space}

Here we concentrate on the case only linear terms of 
$\phic$ are important and multiplicative noise is negligible in the
equation of motion (\ref{eqn:a4eqm}). 
Then the equation of
motion reads
\bea
 (\,\Box+M^2\,)\,\phi_{c}(x)
          + \int_{-\infty}^{t}dt'\int d^3x'C_{a}(x-x')\phi_{c}(x') 
           =\xi_{a}(x) \:. \label{lineareq}
\eea
  Hereafter we omit the
suffix $c$.

In this regime it is convenient to rewrite the above equation in
the wavenumber space.  Defining the spatial Fourier transform as
\bea
 \phik(t)&\equiv&\int d^3x\phi(\vex,t)e^{-i\vecs
 k\cdot \vecs x },~~~\xik(t)
\equiv\int d^3x\xi_a(\vex,t)e^{-i\vecs
 k\cdot \vecs x },\\
 \Cak(t-t')&\equiv& \int d^3xC_a(\vex,t-t')e^{-i\vecs
 k\cdot \vecs x }, 
\eea
we find
\beq
  \ddot\phik(t)+\lmk \vek^2 +M^2\rmk \phik(t) + \int_{-\infty}^t dt'
 \Cak(t-t')\phi_{\mbox{\boldmath\tiny $k$}}(t')=\xik (t) \:,  \label{keq}
\eeq
where $\Cak(t-t')$ is a real function thanks to (\ref{parityeven}).
Here the noise term in the Fourier space, $\xik (t)$, is a random
Gaussian variable with the dispersion,
\beq
  \langle \xik(t)\xi^\ast_{\vecs k'}(t') \rangle 
=\int d^3x D_a(\vex,t-t')e^{-i\vecs k\cdot \vecs x }
(2\pi)^3\delta(\vect k-\vect k')
\equiv D_{a\vecs k}(t-t')
(2\pi)^3\delta(\vect k-\vect k'). \label{xikcorrelation}
\eeq
Thus each Fourier mode is completely decoupled from each other
in the linear regime even
in the presence of the noise term, as
it should be. 

Equation (\ref{keq}) can be solved in terms of
the Fourier transform with respect
to $t$,
\bea
  \tilde\phi(\omega)\equiv\int dt\phi(t) e^{i\omega t},~~~
~&&~~~\tilde\xik(\omega)
\equiv\int dt\xik(t)e^{i\omega t},\\
 \tilde\Cak(\omega)\equiv \int dt\Cak (t)e^{i\omega t},~&&~
\tilde\Dak(\omega)\equiv \int dt\Dak (t)e^{i\omega t}.  \label{Dfourier}
\eea
Here note that $\tilde\Cak(\omega)$ is pure imaginary due to (\ref{timeodd}).
Using the formula 
\beq
  \int_0^\infty d\tau e^{i(\omega-\omega')\tau}
=i{\rm P}\frac{1}{\omega-\omega'}+\pi \delta(\omega-\omega'),
\eeq
we find
\bea
(-\omega^2+\vek^2+M^2)\tilde\phik(\omega)
+\int\frac{d\omega'}{2\pi}{\rm
P}\frac{1}{\omega-\omega'}i\tilde\Cak(\omega')\tilde\phik(\omega)
+\frac{1}{2}\tilde\Cak(\omega)\tilde\phik(\omega)=\tilde\xik(\omega).
\eea
Defining real quantities
\beq
  M_k^2\equiv M^2+\vek^2+\int\frac{d\omega'}{2\pi}{\rm
P}\frac{1}{\omega-\omega'}i\tilde\Cak(\omega'),~~~
\tilde\Gamma_k(\omega)\equiv i\frac{\tilde\Cak(\omega)}{2\omega},
\label{Gammadef}
\eeq
which respectively constitute real and imaginary parts of the self
energy
of $\phi$,
we obtain
\beq
 \phik(t)=-\int\frac{d\omega}{2\pi}\frac{\tilde\xik(\omega)e^{-i\omega t}}
{\omega^2-M_k^2+i\omega\tilde\Gamma_k(\omega)}
=-\int_{-\infty}^{\infty}dt'\int\frac{d\omega}{2\pi}
\frac{\xik(t')e^{i\omega(t'-t)}}{\omega^2-M_k^2+i\omega\tilde\Gamma_k(\omega)}.
\label{feq}
\eeq
If $\tilde\Gamma_k(\omega)$  satisfies
 $0< \tilde\Gamma_k(\omega) \ll M_k$ and $\omega$-dependent part of $M_k$ is
negligibly small, which turn out to be the case
in the specific examples discussed later, (\ref{feq}) has poles at
$\omega\cong \pm M_k-i\tilde\Gamma_k(M_k)/2$ and it can be solved as
\beq
 \phik(t)=\frac{1}{M_k}\int_{-\infty}^t dt' 
e^{-\frac{1}{2}\tilde\Gamma_k(M_k)(t-t')}
 \sin M_k(t-t')\xik(t'). 
\eeq
Adding two independent homogeneous modes, a 
general solution with an arbitrary initial condition $\phik(t_i)$
and $\dot\phik(t_i)$ at some initial time $t=t_i$ is given by
\bea
 \phik(t) &=& \phik(t_i)e^{-\frac{1}{2}\tilde\Gamma_k(M_k)(t-t_i)}\cos M_k(t-t_i)
 + \frac{\dot\phik(t_i)}{M_k}e^{-\frac{1}{2}\tilde\Gamma_k(M_k)(t-t_i)}
\sin M_k(t-t_i) \non\\
 &&+\frac{1}{M_k}\int_{t_i}^t dt' e^{-\frac{1}{2}\tilde\Gamma_k(M_k)(t-t')}
 \sin M_k(t-t')\xik(t'), \label{phisol}
\eea
by virtue of the assumption $\tilde\Gamma_k(M_k) \ll M_k$.

Then using (\ref{xikcorrelation}) and (\ref{Dfourier}),
the expectation value of the absolute square amplitude 
at late time $t \gg t_i +\tilde\Gamma_k^{-1}(M_k)$ reads
\beq
 \la |\phik(t)|^2 \ra=\frac{\tilde\Dak(M_k)}{2M_k^2\tilde\Gamma_k(M_k)}
 \lkk 1+ \frac{\tilde\Gamma_k(M_k)}{M_k}\sin 2M_kt\rkk (2\pi)^3\delta(\vect 0).
 \label{average}
\eeq
The second term in the bracket vanishes of course 
if we take time average over an oscillation period as well.  
Equations (\ref{phisol}) and (\ref{average})
indicate that each mode does not decay completely but its
 square amplitude  approaches an equilibrium value
 determined by the ratio of the power spectrum of the noise 
to $\tilde\Gamma_k(M_k)$ with the time scale 
$\tilde\Gamma_k^{-1}(M_k)=-2iM_k/\tilde\Cak(M_k)$.

In order to evaluate these quantities we must calculate Fourier
transform of
the memory kernel $C_a$ and the noise correlation $D_a$ explicitly
using the expressions given in the previous section. 
Below we study the effects of interactions with fermions and bosons
separately in turn, because they have different behaviors due to the
different statistical properties.  The striking difference between
fermionic noises and bosonic noises have been pointed out in \cite{Yamaguchi:1996dp}.

\subsection{Interaction with fermions}\label{intfermion}

First we study the case the scalar field interacts only with fermions. 
In this case $C_a$ and $D_a$ are governed by $ L_4$, namely (\ref{34})
and (\ref{40}). 
Because both $A_4(x-x')$ and $B_4(x-x')$ are parity-even functions, see
(\ref{parityeven}) and (\ref{parityevenb}), the spatial Fourier
transform is identical to the Fourier cosine transform, so we find
\bea
  C_{a\vecs k}(t-t')=
A_{4\vecs k}(t-t')&=&-2f^2\int d^3x e^{-i\vecs k\cdot\vecs x} 
\mbox{Im}\lnk\tr\lkk S^{F}_{\psi}(\vex,t-t')
S^{F}_{\psi}(-\vex, t'-t)\rkk\rnk  \non\\
&=&-2f^2\mbox{Im}\lnk\tr\lkk S^{F}_{\psi}(\vect p,t-t')
S^{F}_{\psi}(\vect p +\vek, t'-t)\rkk\rnk.
\eea
Here we have used  the fermion propagator expressed by the real time $t$ and
spatial wavenumber $\vep$,
\bea
 S_\psi^F(\vep,t)&=& \int d^3xS_\psi^F(\vect x,t)e^{-i\vecs
 p\cdot \vecs x } \non \\
&=&
\lkk \frac{E_p\gamma_0-\vep\veg+m_\psi}{2E_p}(1-n_p^F)e^{-iE_pt}
-
\frac{-E_p\gamma_0-\vep\veg+m_\psi}{2E_p}n_p^Fe^{iE_pt}\rkk\theta(t)
\label{fermionp} \\
&-&\lkk \frac{E_p\gamma_0-\vep\veg+m_\psi}{2E_p}n^p_Fe^{-iE_pt}
- \frac{-E_p\gamma_0-\vep\veg+m_\psi}{2E_p}(1-n_p^F)e^{iE_pt}\rkk\theta(-t),
\non
\eea
with $E_p\equiv\sqrt{\vep^2+m_\psi^2}$ and $n_{p}^F\equiv n_F(E_p)$.
Using
\bea
\tr\lkk S_\psi^F(\vep,\tau)\right.&&\!\!\!\!\!\!
\left.S_\psi^F(\vep+\vek,-\tau)\rkk
=\non\\
-\frac{1}{E_pE_{k+p}}&&\!\!\!\lkk~~ (E_pE_{k+p}-\vep\cdot\vek-\vep^2+m_\psi^2)
(1-n_{p}^F)n_{k+p}^Fe^{-i(E_p-E_{p+k})\tau}\right. \non \\
&&~~-(-E_pE_{k+p}-\vep\cdot\vek-\vep^2+m_\psi^2)
n_{p}^Fn_{k+p}^Fe^{i(E_p+E_{p+k})\tau} \non \\
&&~~-(-E_pE_{k+p}-\vep\cdot\vek-\vep^2+m_\psi^2)
(1-n_{p}^F)(1-n_{k+p}^F)e^{-i(E_p+E_{p+k})\tau} \non \\
&&~~\left.+(E_pE_{k+p}-\vep\cdot\vek-\vep^2+m_\psi^2)
n_{p}^F(1-n_{k+p}^F)e^{i(E_p-E_{p+k})\tau} \rkk,
\eea
with $\tau\equiv t-t' > 0$, we find that the Fourier transform of the
memory kernel is given by
\bea
&& \tilde C_{a\vecs k}(\omega)
=\int d\tau A_{4\vecs k}(\tau)e^{i\omega\tau}
=-2i\pi f^2\int \frac{d^3p}{(2\pi)^3}
\frac{1}{E_pE_{k+p}} \non\\
&&\times\lnk (E_pE_{k+p}-\vep^2+m_\psi^2)
  \lkk (1-n_{p}^F)n_{k+p}^F - n_{p}^F(1-n_{k+p}^F)\rkk \delta
  (\omega+E_{k+p}-E_p)\right. \non\\
&&~~+(E_pE_{k+p}-\vep^2+m_\psi^2)
  \lkk n_{p}^F(1-n_{k+p}^F) - (1-n_{p}^F)n_{k+p}^F\rkk \delta
  (\omega-E_{k+p}+E_p) \non\\
&&~~+(E_pE_{k+p}+\vep^2-m_\psi^2)
  \lkk n_{p}^Fn_{k+p}^F - (1-n_{p}^F)(1-n_{k+p}^F)\rkk \delta
  (\omega+E_{k+p}+E_p) \non\\
&&~~\left.+(E_pE_{k+p}+\vep^2-m_\psi^2)
  \lkk (1-n_{p}^F)(1-n_{k+p}^F) - n_{p}^Fn_{k+p}^F\rkk \delta
  (\omega-E_{k+p}-E_p)\rnk, \label{fermionC}
\eea
where $E_{k+p}\equiv \sqrt{(\vek+\vep)^2+m_\psi^2}$ 
and $n_{k+p}^F\equiv n_F(E_{k+p})$.
The first term in each square bracket can be interpreted as decay 
or absorption of $\tilde\phik(\omega)$, which is denoted by 
$ R_D$, while
the second term corresponds to inverse decay or creation of 
$\tilde\phik(\omega)$ denoted by
$ R_C$, because $n_q^F$ corresponds to the number density of an
initial state and $1-n_q^F$ to the Pauli-blocking factor of a final state. 
The above expression (\ref{fermionC}) is closely related with the
discontinuity of the self energy of $\phi$ at finite temperature
which was obtained by Weldon \cite{Weldon} using a different
procedure.  In his approach one had to add and subtract appropriate
combinations of $n_p^F$ and $n_{k+p}^F$ to obtain the above form in
which physical interpretation of absorption and creation of $\phi$ is
manifest, while in our scheme the above result is obtained 
straightforwardly from 
the structure of the fermion propagator (\ref{fermionp}).

Due to
the delta function the ratio of creation and destruction rates
satisfies the detailed-balance relation,
\beq
\frac{ R_C}{ R_D}=e^{-\beta\omega},  \label{balance}
\eeq
for all combinations.
For example, in the first square bracket of the right-hand-side of
(\ref{fermionC}), we find
\beq
 \frac{ R_C}{ R_D}
=\frac{n_{p}^F(1-n_{k+p}^F)}{(1-n_{p}^F)n_{k+p}^F}
=e^{\beta(E_{k+p}-E_{p})}=e^{-\beta\omega},
\eeq
under the condition $\omega=E_p-E_{k+p}$ coming from the delta function
$\delta(\omega+E_{k+p}-E_p)$.

The dispersion of the stochastic noise in Fourier space, on the
other hand, reads
\bea
&& \tilde D_{a\vecs k}(\omega)=\int d\tau B_{4\vecs
 k}(\tau)e^{i\omega\tau}
=\pi f^2\int \frac{d^3p}{(2\pi)^3}\frac{1}{E_pE_{k+p}}
\non \\
&&\times\lnk (E_pE_{k+p}-\vep^2+m_\psi^2)
  \lkk (1-n_{p}^F)n_{k+p}^F + n_{p}^F(1-n_{k+p}^F)\rkk \delta
  (\omega+E_{k+p}-E_p)\right. \non\\
&&~~+(E_pE_{k+p}-\vep^2+m_\psi^2)
  \lkk n_{p}^F(1-n_{k+p}^F) + (1-n_{p}^F)n_{k+p}^F\rkk \delta
  (\omega-E_{k+p}+E_p) \non\\
&&~~+(E_pE_{k+p}+\vep^2-m_\psi^2)
  \lkk n_{p}^Fn_{k+p}^F + (1-n_{p}^F)(1-n_{k+p}^F)\rkk \delta
  (\omega+E_{k+p}+E_p) \non\\
&&~~\left.+(E_pE_{k+p}+\vep^2-m_\psi^2)
  \lkk (1-n_{p}^F)(1-n_{k+p}^F) + n_{p}^Fn_{k+p}^F\rkk \delta
  (\omega-E_{k+p}-E_p)\rnk, \label{fermionD}
\eea
In this dispersion, both destruction $ R_D$ and creation $ R_C$
contribute additive manner.
From (\ref{Gammadef}), (\ref{fermionC}) and (\ref{fermionD}), we find
\beq
 \frac{\tilde\Dak(\omega)}{\tilde\Gamma_k(\omega)}
=-2i\omega\frac{\tilde\Dak(\omega)}{\tilde\Cak(\omega)}
=\omega\frac{ R_D+ R_C}{ R_D- R_C}
=\omega\frac{e^{\beta\omega}+1}{e^{\beta\omega}-1}\cong
2T,  \label{DGamma}
\eeq
where the last approximate equality holds for the soft modes with
$\omega \ll T$.  This is nothing but the
fluctuation-dissipation relation derived purely from
quasi-nonequilibrium quantum field theory at finite temperature.

Note that the fluctuation-dissipation relation has also been
obtained by Gleiser and Ramos \cite{GR} in the context of nonequilibrium
field theory at finite temperature.  However, because they assumed the
scalar field evolves adiabatically, they had to invoke higher loop
effects to obtain a  nonvanishing dissipation coefficient.  As a result 
their noise term and dissipation term appear at different order of
perturbation.  
This problem of adiabatic treatment has been pointed out  by Gleiner and
M\"uller \cite{GM} who adopted a harmonic approximation instead in order
to extract a term proportional to $\dot\phi$, which represents
dissipation
 in the equation of motion and obtained the correct result.
In the present analysis we have made no assumption about
the adiabaticity of the evolution of the scalar field but worked in the
Fourier space assuming that the quartic term dominates its potential.
Then we can see that both dissipation and noise terms appear at the
same order of perturbation.  

The time average of (\ref{average}) over an
oscillation period reads
\beq
  \frac{1}{2}\langle |\dot\phik(t)|^2\rangle\cong
\frac{1}{2}M_k^2\langle |\phik(t)|^2\rangle =
\frac{1}{2}T(2\pi)^3\delta(\vect 0).
\eeq
This equation shows the classical 
equipartition law is satisfied for low-momentum modes that kinetic
energy per degree of freedom is equal to $T/2$.  This property can be seen
more manifestly if we adopt a box normalization with a finite side $L$
and periodic boundary condition.  Then $\phi(\vex,t)$ is expanded as
\beq
  \phi(\vex,t)=\sum_{\vecs n} \phi_{\vecs n}(t)
e^{i\frac{2\pi}{L}{\vecs n\cdot\vecs x}},
\eeq
with $\vect n$ being a spatial vector consisting of integers.
Then (\ref{xikcorrelation}) is replaced by
\beq
  \langle \xi_{\vecs n}(t)\xi^\ast_{\vecs n'}(t') \rangle 
=D_{a \vecs n}(t-t')\delta_{\vecs n \vecs n'},~~~
D_{a \vecs n}(t-t')\equiv \int_0^L\frac{d^3x}{L^3}
D_a(\vex,t-t')e^{-i\frac{2\pi}{L}{\vecs n\cdot\vecs x}}.
\eeq
From (\ref{average}) we find
average kinetic energy of each soft mode is given by
\beq
  \frac{1}{2}\langle |\dot\phi_{\vecs n}(t)|^2\rangle
  =\frac{1}{2}T.  \label{equipartition}
\eeq

The above is the results for the Rayleigh-Jeans regime
$\omega \ll T$, where classical
analysis applies.  We now consider a more general case.  Instead of taking
the high temperature limit $\omega/T\longrightarrow 0 $ as in the last
equality of (\ref{DGamma}), we rewrite (\ref{DGamma}) as
\beq
 \frac{\tilde\Dak(\omega)}{\tilde\Gamma_k(\omega)}
=\omega\frac{ R_D+ R_C}{ R_D- R_C}
=\omega\frac{e^{\beta\omega}+1}{e^{\beta\omega}-1}
=2\omega\lmk n_B(\omega)+\frac{1}{2}\rmk.
\eeq
Then (\ref{average}) reads
\beq
 \frac{M_k^2\la |\phik(t)|^2 \ra}{V}=
\frac{\tilde\Dak(M_k)}{\tilde\Gamma_k(M_k)V}
=M_k\lmk n_B(M_k)+\frac{1}{2}\rmk,
\eeq
where $V\equiv (2\pi)^3\delta(\vect 0)$ denotes (infinite) spatial 
volume.  Its interpretation is obvious.  The left-hand-side represents
energy density stored in the $\vek$-mode and the right-hand-side shows it
consists of thermal and vacuum quanta with energy level $M_k$ 
in the final equilibrium state.
Thus the interaction with a thermal bath drives each Fourier mode $\phik(t)$
to the thermal equilibrium value with the same temperature in the time
scale $\tilde\Gamma_k(M_k)^{-1}$.

Next we evaluate the dissipation rate using (\ref{fermionC}).  Since we
are primarily interested in the fate of the homogeneous coherent mode, 
we take $\vek =\vect 0$.  Then only the last term of (\ref{fermionC}) is
nonvanishing and we find
\beq
  \Gamma_F(T)\equiv\tilde\Gamma_0(M_0)
=\frac{f^2}{8\pi}M_0\lkk 1- 
\lmk\frac{2m_\psi}{M_0}\rmk^2\rkk^{3/2}
\lkk 1-2n_F\lmk\frac{M_0}{2}\rmk\rkk =\Gamma_F(0) \lkk
1-2n_F\lmk\frac{M_0}{2}\rmk\rkk. \label{fermiondissipate}
\eeq
We thus find the dissipation rate at finite temperature is suppressed by 
the last factor in (\ref{fermiondissipate}) due to Pauli blocking with
$\Gamma_F(0)$ 
being the decay rate of a $\phi$ particle at rest
into two fermions $\psi$ and $\bar\psi$ at zero temperature.
Note that the above dissipation rate vanishes when $M_0 < 2m_\psi$.  In this
case the coherent oscillation is not thermalized through the
Yukawa interaction at one-loop, because not only the dissipation kernel 
$ \tilde{C}_{a\vecs 0}(\omega)$ but also noise correlation
$ \tilde{D}_{a\vecs 0}(\omega)$ vanishes in this case since both contain
delta functions with the same arguments.

The above arguments are based on the propagator (\ref{fermionp}) where 
only the zero-temperature intrinsic mass $m_\psi$ is taken into
account.  If the Yukawa interaction
$f\phi\overline\psi \psi$ generates large oscillating mass to $\psi$, 
decay of $\phi$ into two fermions would be possible only during a short
interval when $f|\phi| < M_0/2$  as the scalar field passes through the
origin twice in each oscillation period.  The dissipation rate of the
scalar field in such a situation cannot be dealt with the perturbation
theory we are using. This issue has been investigated by Dolgov
and Kirilova \cite{Dolgov:1989us} using a quasiclassical approximation
at zero temperature.  They find that the dissipation rate of $\phi$ is
not exponentially suppressed but by a factor $\sim (M_0/m_{\chi,\rm
osc})^{1/2}$, where $m_{\chi,\rm osc}$ is the maximum of $\chi$'s mass with
the oscillating component taken into account.  Finite-temperature
generalization of the analysis in such a regime is not straightforward
and we restrict our analysis to the perturbative regime 
$f|\phi| \lesssim M_0$
here.  

On the other hand, recently Kolb, Notari, and Riotto  \cite{Kolb:2003ke}
 argue that if the would-be decay products of the oscillating
inflaton scalar field acquire a thermal mass larger than the inflaton
mass in the thermal background, the inflaton cannot decay into these
particles, and that reheating is suspended for some time based on the
observation that the phase space would be closed for the mass of the
decay product being larger than half the inflaton mass.  This phenomenon
could be observed in our perturbative approach as well, if we used,
instead of finite-temperature ``bare'' propagator (\ref{fermionp}), the
``dressed'' propagator in which finite-temperature higher-order
quantum corrections
are taken into account.  But use of such an dressed propagator can
easily result in overcounting of diagrams and the detailed comparison of
two different expansion method is still under way.  Here we focus on
the effects from  lowest possible orders and continue to use the bare
propagators.  In the practical applications in \S \ref{inflation} and
\S \ref{flatdirection} we mostly consider the cases
the thermal mass of decay products remain smaller than the angular
frequency of the oscillating field, so both approaches give the same
results.

\subsection{Interaction with bosons}
\label{threebody}

Next we consider the effect of three-body interaction $\CM\phi\chi^2$
for which $C_a$ and $D_a$ are determined by $ L_5$, namely (\ref{35})
and (\ref{41}).
Again their spatial Fourier transform is identical to the Fourier cosine
transform due to the parity evenness, and the memory kernel reads
\bea
 C_{a\vecs k}(\tau)=A_{5\vecs k}(\tau)&=&
4\CM^2 \int d^3x e^{-i\vecs k\cdot\vecs x}\im\lkk
G_\chi^F(\vex,\tau)^2\rkk
\non \\
&=&4\CM^2 \int \frac{d^3p}{(2\pi)^3}\im\lkk G_\chi^F(\vep,\tau)
G_\chi^F(\vek-\vep,\tau)\rkk,
\eea
for $\tau=t-t' > 0$ and $C_{a\vecs k}(\tau)=-C_{a\vecs k}(-\tau)$ for
$\tau < 0$.
Here $ G_\chi^F(\vep,\tau)$ is defined by
\bea
 G_\chi^F(\vep,\tau)&=&\int d^3xG_\chi^F(\vect x,\tau)e^{-i\vecs
 p\cdot \vecs x } \non \\
&=&\frac{1}{2\omp}\lnk \lkk 1+n_B(\omp)\rkk e^{-i\omp |\tau|}
+n_B(\omp)e^{i\omp |\tau|}\rnk , ~~~ \omp\equiv \sqrt{\vep^2+m_\chi^2}.
\eea
We obtain
\bea
 \tilde{C}_{a\vecs k}(\omega)
&=&-i\pi\CM^2\int\frac{d^3p}{(2\pi)^3}
\frac{1}{\omega_p\omega_{k-p}} \non\\
&&~~~~~~~~\times\lnk
\lkk (1+n_p)(1+n_{k-p})-n_pn_{k-p}\rkk
\delta(\omega-\omega_p-\omega_{k-p})\right. \non \\
&&~~~~~~~~~~
+\lkk (1+n_p)n_{k-p}-(1+n_{k-p})n_p\rkk
\delta(\omega-\omega_p+\omega_{k-p})   \non \\
&&~~~~~~~~~~+\lkk n_{p}(1+n_{k-p})-(1+n_{p})n_{k-p}\rkk
\delta(\omega+\omega_p-\omega_{k-p}) \non \\
&&~~~~~~~~~~\left.
+\lkk n_pn_{k-p}-(1+n_p)(1+n_{k-p})\rkk
\delta(\omega+\omega_p+\omega_{k-p})\rnk,  \label{Cboson}
\eea
where $\omega_{k-p}\equiv\sqrt{(\vek-\vep)^2+m_\chi^2}$, 
$n_p\equiv n_B(\omega_p)$, 
and $n_{k-p}\equiv n_B(\omega_{k-p})$, respectively.  
The first term in each bracket represents destruction ($ R_D$) while
the second term corresponds to creation 
($ R_C$).  
Due to
the delta function their ratio satisfies the detailed-balance relation,
$ R_C/ R_D=e^{-\beta\omega}$, for all combinations.
We can also confirm that $\tilde\Gamma_k(\omega)=i\tilde{C}_{a\vecs
k}(\omega)/2\omega$
is positive definite.

Next we consider the power spectrum of thermal noise given by
\beq
\Dak(\tau)=B_{5\vecs k}(\tau)=2\CM^2\int\frac{d^3p}{(2\pi)^3}
\re\lkk G_\chi^F(\vep,\tau)G_\chi^F(\vek-\vep,\tau)\rkk.
\eeq
Its Fourier transform with respect $\tau$ reads
\bea
\tilde\Dak(\omega)&=&\frac{\pi}{2}\CM^2\int\frac{d^3p}{(2\pi)^3}
\frac{1}{\omega_p\omega_{k-p}} \non \\
&&\times\lnk
\lkk (1+n_p)(1+n_{k-p})+n_pn_{k-p}\rkk
\delta(\omega-\omega_p-\omega_{k-p})\right. \non \\
&&
+\lkk (1+n_p)n_{k-p}+(1+n_{k-p})n_p\rkk
\delta(\omega-\omega_p+\omega_{k-p})   \non \\
&&+\lkk n_{p}(1+n_{k-p})+(1+n_{p})n_{k-p}\rkk
\delta(\omega+\omega_p-\omega_{k-p}) \non \\
&&\left.
+\lkk n_pn_{k-p}+(1+n_p)(1+n_{k-p})\rkk
\delta(\omega+\omega_p+\omega_{k-p})\rnk, \label{Dboson}
\eea
As in the case of Yukawa interaction (\ref{fermionD}), delta functions in 
(\ref{Dboson}) have been multiplied by $ R_D+ R_C$, which means that
both destruction and creation act as a noise to the evolution of the
scalar field in the same way.

From (\ref{Cboson}) and (\ref{Dboson}) we find again that
\beq
 \frac{\tilde\Dak(\omega)}{\tilde\Gamma_k(\omega)}
=\omega\frac{ R_D+ R_C}{ R_D- R_C}
=\omega\frac{e^{\beta\omega}+1}{e^{\beta\omega}-1}
=2\omega\lmk n_B(\omega)+\frac{1}{2}\rmk, \label{DGammab}
\eeq
namely,
\beq
 \frac{\tilde D_{a\vecs k}(M_k)}{\tilde\Gamma_k(M_k)}=2T,  \label{2t}
\eeq
in the Rayleigh-Jeans limit $M_k \ll T$. Thus
 the fluctuation-dissipation theorem is satisfied in this case, too,
 and the final equilibrium configuration has the same property as in the
 case thermalization proceeds through Yukawa interaction.

As explained in the previous section, 
the dissipation rate toward thermal equilibrium distribution is
given by $\tilde\Gamma_k(M_k)$.  For the coherent zero-mode, in which we
are primarily interested, one can easily find 
\beq
  \Gamma_B(T)\equiv\tilde\Gamma_0(M_0)=\frac{\CM^2}{8\pi M_0}
\lkk 1-\lmk\frac{2m_\chi}{M_0}\rmk^2\rkk^{1/2}
\lkk 1+2n_B\lmk\frac{M_0}{2}\rmk \rkk
=\Gamma_B(0)
\lkk 1+2n_B\lmk\frac{M_0}{2}\rmk \rkk, \label{chidecay}
\eeq
because only the first $\delta$ function in (\ref{Cboson}) gives
nonvanishing contribution when $\vek = \vect 0$.
Again $\Gamma_B(0)$ 
is the decay rate of  $\phi$ into two $\chi$
particles through trilinear interaction.
Thus the dissipation rate is enhanced at finite temperature 
due to the presence of bosons. In the high temperature limit $\beta M_0
\ll 1$ (\ref{chidecay}) reads 
\beq
 \Gamma_B(T)\simeq \frac{4T}{M_0}\Gamma_B(0)
 =\frac{\CM^2T}{2\pi M_0^2}
\lkk 1-\lmk\frac{2m_\chi}{M_0}\rmk^2\rkk^{1/2}. \label{chidecayhigh}
\eeq
Note that these dissipation rates vanish when $M_0 < 2m_\chi$.  In this
case the coherent oscillation is not thermalized through one-loop of
the three body
interaction $\CM \phi\chi^2$.  For the same reason described in the
latter part of \S \ref{intfermion}, the above dissipation rate applies only
for $M_0^2 \gtrsim \CM\phi$.  In the large field-amplitude regime when
this inequality is not satisfied, particle creation through broad
parametric resonance would be much more efficient \cite{Kofman:1994rk}.

\subsection{Setting-sun diagrams}

Next we study the contribution from the setting-sun diagrams, $ L_2$
and $ L_7$.  Since $ L_7$ is expected to give larger
contribution we first analyze the 
 Fourier transform of the memory kernel corresponding to it, which 
is given by
\bea
 \tilde\Cak(\omega)&=&\int dt A_{7\vecs k}(\tau)e^{i\omega \tau} \non\\
 &=&h^4\int d\tau e^{i\omega \tau}\int \frac{d^9p}{(2\pi)^9}(2\pi)^3
 \delta (\vep_1+\vep_2+\vep_3-\vek)\im\lkk G_\chi^F(\vep_1,\tau)
 G_\chi^F(\vep_2,\tau)G_\phi^F(\vep_3,\tau)\rkk \non\\
 &=&-i\pi h^4\int \frac{d^9p}{(2\pi)^9}(2\pi)^3
 \delta (\vep_1+\vep_2+\vep_3-\vek)\frac{1}{8\omega_1\omega_2\omega_3} \non\\
 &&\times\lnk~\lkk (1+n_1)(1+n_2)(1+n_3)-n_1n_2n_3\rkk
\delta(\omega-\omega_1-\omega_2-\omega_3)\right. \non\\
 &&~~~+\lkk n_1n_2n_3-(1+n_1)(1+n_2)(1+n_3)\rkk
\delta(\omega+\omega_1+\omega_2+\omega_3) \non\\
&&~~~+\lkk n_1(1+n_2)(1+n_3)-(1+n_1)n_2n_3\rkk
\delta(\omega+\omega_1-\omega_2-\omega_3) \non\\
 &&~~~+\lkk (1+n_1)n_2n_3-n_1(1+n_2)(1+n_3)\rkk
\delta(\omega-\omega_1+\omega_2+\omega_3) \non\\
&&~~~+\lkk (1+n_1)n_2(1+n_3)-n_1(1+n_2)n_3\rkk
\delta(\omega-\omega_1+\omega_2-\omega_3) \non\\
&&~~~+\lkk n_1(1+n_2)n_3-(1+n_1)n_2(1+n_3)\rkk
\delta(\omega+\omega_1-\omega_2+\omega_3) \non\\
&&~~~+\lkk n_1n_2(1+n_3)-(1+n_1)(1+n_2)n_3\rkk
\delta(\omega+\omega_1+\omega_2-\omega_3) \non\\
&&~~~\left.+\lkk (1+n_1)(1+n_2)n_3-n_1n_2(1+n_3)\rkk
\delta(\omega-\omega_1-\omega_2+\omega_3)\rnk . \label{settingsunC}
\eea
Here we have defined $\omega_1\equiv \sqrt{\vep_1^2+m_\chi^2}$,
$\omega_2\equiv \sqrt{\vep_2^2+m_\chi^2}$,  $\omega_3\equiv
\sqrt{\vep_3^2+m_\phi^2}$, and $n_i\equiv n_B(\omega_i)$.
Once again the first term in each coefficient of delta functions
 represents destruction ($ R_D$) while
the second term corresponds to creation 
($ R_C$).  
Due to
the delta function their ratio satisfies the detailed-balance relation,
$ R_C/ R_D=e^{-\beta\omega}$, for all combinations as before.
We can also confirm that $\tilde\Gamma_k(\omega)=i\tilde{C}_{a\vecs
k}(\omega)/2\omega$
is positive definite.

Similarly, the Fourier transform of noise correlation reads
\bea
 \tilde\Dak(\omega)&=&\int dt B_{7\vecs k}(\tau)e^{i\omega t} \non\\
 &=&\frac{h^4}{2}\int dte^{i\omega t}\int \frac{d^9p}{(2\pi)^9}(2\pi)^3
 \delta (\vep_1+\vep_2+\vep_3-\vek)\re\lkk G_\chi^F(\vep_1,\tau)
 G_\chi^F(\vep_2,\tau)G_\phi^F(\vep_3,\tau)\rkk \non\\
 &=&\frac{\pi h^4}{2}\int \frac{d^9p}{(2\pi)^9}(2\pi)^3
 \delta (\vep_1+\vep_2+\vep_3-\vek)\frac{1}{8\omega_1\omega_2\omega_3} \non\\
 &&\times\lnk~\lkk (1+n_1)(1+n_2)(1+n_3)+n_1n_2n_3\rkk
\delta(\omega-\omega_1-\omega_2-\omega_3)\right. \non\\
 &&~~~+\lkk n_1n_2n_3+(1+n_1)(1+n_2)(1+n_3)\rkk
\delta(\omega+\omega_1+\omega_2+\omega_3) \non\\
&&~~~+\lkk n_1(1+n_2)(1+n_3)+(1+n_1)n_2n_3\rkk
\delta(\omega+\omega_1-\omega_2-\omega_3) \non\\
 &&~~~+\lkk (1+n_1)n_2n_3+n_1(1+n_2)(1+n_3)\rkk
\delta(\omega-\omega_1+\omega_2+\omega_3) \non\\
&&~~~+\lkk (1+n_1)n_2(1+n_3)+n_1(1+n_2)n_3\rkk
\delta(\omega-\omega_1+\omega_2-\omega_3) \non\\
&&~~~+\lkk n_1(1+n_2)n_3+(1+n_1)n_2(1+n_3)\rkk
\delta(\omega+\omega_1-\omega_2+\omega_3) \non\\
&&~~~+\lkk n_1n_2(1+n_3)+(1+n_1)(1+n_2)n_3\rkk
\delta(\omega+\omega_1+\omega_2-\omega_3) \non\\
&&~~~\left.+\lkk (1+n_1)(1+n_2)n_3+n_1n_2(1+n_3)\rkk
\delta(\omega-\omega_1-\omega_2+\omega_3)\rnk .
\label{settingsunD}
\eea
Each coefficient of delta functions consists of $R_D+R_C$ as before.
One can also calculate the respective quantities for the other setting-sun
diagram $ L_2$, which has also been calculated in \cite{GM},
 by the following
replacement:
\bea
&&h^4\longrightarrow \lambda^2/3, \non \\
&&G_\chi^F(\vep_1,\tau)\longrightarrow G_\phi^F(\vep_1,\tau),~~  
G_\chi^F(\vep_2,\tau)\longrightarrow G_\phi^F(\vep_2,\tau), \label{replace}  \\ 
&&\omega_1\longrightarrow
\sqrt{\vep_1^2+m_\phi^2}, ~~  \omega_2\longrightarrow
\sqrt{\vep_2^2+m_\phi^2}, ~~{\rm with}~ 
n_i= n_B(\omega_i). \non
\eea

In both cases we  find the same structure again
for $\tilde\Dak(\omega)$ and 
$\tilde\Gamma_k(\omega)=i\tilde\Cak(\omega)/2\omega$, that is,
\beq
 \frac{\tilde\Dak(\omega)}{\tilde\Gamma_k(\omega)}
=\omega\frac{ R_D+ R_C}{ R_D- R_C}
=\omega\frac{e^{\beta\omega}+1}{e^{\beta\omega}-1}
=2\omega\lmk n_B(\omega)+\frac{1}{2}\rmk,  \label{DGammas}
\eeq
and the fluctuation-dissipation relation is satisfied.

Since the setting-sun diagrams involve three particles in the
intermediate states  their contribution for the zero mode 
is nonvanishing even when $M_0/3$ is smaller than the mass of the
interchanged particles.  Hence this two-loop effect could be important 
when $M_0$ is smaller than $2m_\chi$ or $2m_\psi$ so that one-loop effects
discussed in the previous sections are inoperative.

The analytic evaluation of the dissipation rate with these diagram
is  cumbersome for general cases, so we report the result only for several
limiting cases.  First in the high temperature limit with
$ T \gg M_0 \gg m_\phi,~m_\chi$, it reads
\beq
 \tilde\Gamma_0(M_0) \simeq \frac{\lambda^2T^2}{192\pi M_0},
 \label{sunsetlambda} 
\eeq
for $\lambda\phi^4/4!$ interaction, and
\beq
 \tilde\Gamma_0(M_0) \simeq \frac{h^4T^2}{64\pi M_0},  \label{sunsetgamma}
\eeq
for $h^2\phi^2\chi^2/4$ interaction, where we have used a formula
\beq
\int_0^1 dx\frac{\ln x}{x^2-1}=\frac{\pi^2}{8},
\eeq  
and neglected $m_\phi$ and $m_\chi$ in the intermediate state.  
In this case, the first, the third, the fifth, and
the last delta functions  of (\ref{settingsunC})
and (\ref{settingsunD}) give nonvanishing contribution.

Since  $h^4$ is likely to be of order of $\lambda ~(\gg \lambda^2)$,
we expect (\ref{sunsetgamma}) is much larger than (\ref{sunsetlambda}),
so we concentrate on the diagram $ L_7$ with $h^2\phi^2\chi^2/4$
interaction from now on and take the masses of interchanged particles
into account.
Then we find 
\beq
 \tilde\Gamma_0(M_0) \simeq \frac{3h^4T^2}{256\pi M_0},  \label{sunsetgammam}
\eeq
for $T \gg M_0=m_\phi \gg m_\chi$.  In this case  
the third, the fifth, and
the last delta functions of (\ref{settingsunC})
and (\ref{settingsunD})
give nonvanishing contribution.
Even in the case the mass of interchanged particle
$m_\chi$ is much heavier than $M_0$ one can easily see that 
$\delta(M_0-\omega_1+\omega_2-\omega_3)$ and
$\delta(M_0+\omega_1-\omega_2-\omega_3)$ 
in (\ref{settingsunC}) and (\ref{settingsunD})
can give nonvanishing
contributions because the large masses in
$\omega_1$ and $\omega_2$ tend to cancel each other in these delta
functions.  
As a result we find that, contrary to the case of Yukawa coupling and
three-body bosonic interaction, 
the dissipation rate due to the setting-sun diagram
is nonvanishing even if $T> m_\chi \gg M_0 \geq m_\phi$, and reads
\beq
   \tilde\Gamma_0(M_0) \simeq \frac{h^4T^2}{128\pi^2m_\chi}.
  \label{Gammassb}
\eeq
Thus it is suppressed only by a factor
$M_0/m_\chi$.  This is the case $m_\chi$ is large and constant.  We have
not manipulated the case $m_\chi$ has a large oscillating component, but
 the suppression might be even milder then.

\subsection{Summary of this section}

Here we summarize the results of our analysis for the case $\phi$ obeys
linear equation of motion with an additive stochastic noise term
(\ref{lineareq}).  Working in the Fourier space we have solved the
equation and obtained a general solution (\ref{phisol})
whose damping rate is proportional to the imaginary part of the Fourier
transform of the memory kernel, $\tilde\Cak(\omega)$, related to 
the self energy.  

We have then shown that the expectation value of the square amplitude of
each wavenumber mode relaxes to a specific value determined by the ratio
of the Fourier transform of the dispersion of noise correlation to the
dissipation rate.  In the Rayleigh-Jeans limit this ratio reduces to the
temperature and the classical fluctuation-dissipation theorem holds
there.
In more general cases we find that the energy density of each mode is
the sum of thermal and zero-point vacuum contributions in the final equilibrium
state.  These results are entirely due to the detailed balance relation
(\ref{balance}) and is independent of the nature of the intermediate
state in the loop diagram.

On the other hand, the high-temperature behaviors
 of the dissipation rate is totally
different whether it arises from fermionic interaction or bosonic
interaction.  Although it is equal to the decay width of $\phi$
particle at zero temperature in the perturbative regime, 
at finite temperature it is suppressed 
in the former case due to Pauli blocking (\ref{fermiondissipate}) 
and enhanced in the latter case
due to the induced effects (\ref{chidecay}).  These properties have also
 been obtained using a different technique \cite{Boyanovsky:1995em}.
  In both cases one-loop
effects are shut off for relaxation of the zero-mode field oscillation 
when $M_0$ is smaller than $2m_\chi$ or $2m_\psi$.  The two-loop diagram
would be very important in such a situation.  The dissipation rate due to
 two-loop setting-sun diagram $L_7$ is summarized as
\bea
  \Gamma_S(T)\cong \lnk\begin{array}{ll}
      \displaystyle{\frac{h^4T^2}{64\pi M_0}}
 & \mbox{for}~T\gg M_0 \gg m_\phi,~m_\chi, \\[5mm]
           \displaystyle{\frac{3h^4T^2}{256\pi M_0}} & \mbox{for}~
 T\gg M_0 = m_\phi \gg m_\chi, \\[5mm]
  \displaystyle{\frac{h^4T^2}{128\pi^2 M_0}}
 & \mbox{for}~ T > m_\chi \gg M_0 \geq  m_\phi.
 \end{array}\right.  \label{settingsungamma}
\eea

\section{Nonlinear regime: Effects of multiplicative noise and dissipation}

So far we have analyzed the linear regime when analytic solution can be
found for each Fourier mode.  Next we step forward to include nonlinear
interactions and analyze the effects of multiplicative noise and
dissipation.  To do this we have to deal with the full equation of
motion (\ref{eqn:a4eqm}) which is not soluble analytically.
Hence we can at best hope to extract a term representing dissipation in
the equation of motion and compare it with the noise correlation.

In the previous case of the linear equation of motion we were able to
find dissipation rate without explicitly extracting a term proportional to 
$\dot\phi$, which typically represents dissipation, unlike previous
literatures \cite{GR,GM}, because we have solved the equation of motion
analytically and read off the dissipation rate from the solution.
Alternatively, however, we may also extract a dissipation term with the
correct magnitude in the equation of motion without knowing a
solution.  Here we first describe such a procedure for the linear equation
of motion as a practice to treat multiplicative noise and dissipation.

\subsection{Alternative
derivation of a dissipation term in the linear equation of motion}
\label{prepare}

Here we return to the linear equation of motion (\ref{lineareq}) and
denote the primitive function of $C_a(\vex,t)$ with respect to $t$  by
$E_a(\vex,t)$, namely,
\beq
  C_a(\vex,t)=\frac{\partial}{\partial t}E_a(\vex,t), \label{Mdef}
\eeq
or
\beq
 C_a(\vex-\vex',t-t')=- \frac{\partial}{\partial t'}E_a(\vex-\vex',t-t').
\eeq
Then after integration by parts with respect to $t'$,
Eq.\ (\ref{lineareq}) reads
\beq
 (\,\Box+M^2\,)\,\phi(x)
  -    \int d^3x' E_a(\vex-\vex',0)\phi(\vex',t)
      + \int_{-\infty}^{t}dt'\int d^3x'E_{a}(x-x')\dot\phi(x') 
           =\xi_{a}(x) \:,  \label{neweq}
\eeq
where we have neglected a contribution from infinite past.
It is evident from  (\ref{Gammadef}) and (\ref{Mdef})
 that the Fourier transform of the new kernel,
$\tilde E_{a\vecs k}(\omega)$, is related to $\tilde\Gamma_{k}(\omega)$
as
\beq
  \tilde E_{a\vecs k}(\omega)=2\tilde\Gamma_{k}(\omega).
\eeq
Let us consider spatially 
 homogeneous field configuration $\phi(\vex,t)=\phi(t)$ or $\vek=\vect
 0$ mode, for which (\ref{neweq}) reads
\beq
 \ddot\phi(t)+M^2\phi(t)-E_{a\vecs 0}(t=0)\phi(t)
+\int_0^\infty E_{a\vecs 0}(\tau)\dot\phi(t-\tau)d\tau=\xi_{\vecs 0}(t),
\label{homoeq}
\eeq
where $E_{a\vecs 0}(\tau)$ and $\xi_{\vecs 0}(t)$ are spatial Fourier
 transform of the respective quantities with $\vek =\vect 0$.

Although the last term of the left-hand-side of (\ref{homoeq})
represents dissipation formally, its magnitude depends how the scalar
field evolves.
For example, if we employ an adiabatic approximation such as
$\phi(t)=\phi(t_i)+\dot\phi(t_i)(t-t_i)$ at this stage, 
$\dot\phi(t-\tau)$ should be treated as a constant in the integrand in
(\ref{homoeq}).  In this case the dissipation term vanishes,
\beq
  \int_0^\infty E_{a\vecs 0}(\tau)\dot\phi(t-\tau)d\tau
  =\int_0^\infty E_{a\vecs 0}(\tau)d\tau\dot\phi
 =\frac{1}{2}\tilde E_{a\vecs 0}(\omega=0)\dot\phi=
 \tilde\Gamma_0(\omega=0)\dot\phi=0.
\eeq
This is the very reason the previous approach in the literatures
 \cite{Mor,GR} had to
invoke higher-loop effects or adopt a different method \cite{Hosoya:1983ke}
to yield a nonvanishing dissipation rate in the adiabatic regime.
On the other hand, if we adopt harmonic expansion around $t$,
\beq
 \phi(t-\tau)=\phi(t)\cos M\tau -\frac{\dot\phi(t)}{M}\sin M\tau,
 \label{harmonic}
\eeq
as was done by Gleiner and M\"uller \cite{GM}, the term proportional to
$\dot\phi(t)$ reads
\beq
  \int_0^\infty E_{a\vecs 0}(\tau)\dot\phi(t-\tau)d\tau
 \supset \int_0^\infty E_{a\vecs 0}(\tau)\cos M\tau d\tau\dot\phi(t)
 = \frac{1}{2}\tilde E_{a\vecs 0}(M)\dot\phi(t)=\tilde\Gamma_0(M)\dot\phi(t),
\label{dispgm}
\eeq
which agrees with our  result in the preceding section
that has been obtained more
straightforwardly.
Thus we can see that the Fourier transform of the new kernel $E_a(x)$
gives the dissipation rate for oscillating fields even if we did not know
its exact solution.

\subsection{fluctuation dissipation theorem for multiplicative noise and
  dissipation}

We now apply the above observation for the full nonlinear evolution
equation (\ref{eqn:a4eqm}) without solving it.
Since we have fully clarified the roles of additive noise and the
corresponding kernel $C_a(x-x')$ in the equation of motion in \S
\ref{linearsection}, 
we omit these terms and consider the following equation.
\bea
  (\,\Box+M^2\,)\,\phi(x)
          +\frac{\lambda}{3!}\phi^3(x) 
+\phi(x)\int_{-\infty}^{t}dt'\int d^3x'
              C_{m}(x-x')\phi^2(x') 
           =\phi(x)\xi_{m}(x) \:.
  \label{multieq}
\eea
In order to follow the same procedure as in \S \ref{prepare}
we define the primitive function of $C_m(\vex,t)$, $E_m(\vex,t)$, in
terms of
\beq
   C_m(\vex,t)=\frac{\partial}{\partial t}E_m(\vex,t),
\eeq
and perform integration by parts with respect to $t'$, to yield
\bea
&&  (\,\Box+M^2\,)\,\phi(x)
          +\frac{\lambda}{3!}\phi^3(x) 
- \int d^3x'\phi(x)E_m(\vex-\vex',0)\phi^2(\vex',t)\non\\
&&\qquad\qquad+\int_{-\infty}^{t}dt'\int d^3x'
             2\phi(x) E_{m}(x-x')\phi(x')
\dot\phi(x') 
           =\phi(x)\xi_{m}(x).  \label{multieq2}
\eea
Here the last term in the left-hand-side
includes effects of dissipation.  So in order to see if the
fluctuation-dissipation relation also holds for the case of multiplicative
noise, we perform the Fourier transform of $\CE(x,x')\equiv
2\phi(x) E_{m}(x-x')\phi(x') $ as
\bea
  \tilde\CE_{\vecs k, \vecs k'}(\omega,\omega')
 &\equiv& \int d^3xdt\int d^3x'dt'
  \CE(x,x')e^{-i\vecs k\cdot \vecs x +i\omega t}
 e^{-i\vecs k'\cdot \vecs x' +i\omega' t'} \non\\
 &=&\int \frac{d^3K}{(2\pi)^3}\frac{d\Omega}{2\pi}
 2\tilde\phi_{\vecs k-\vecs K}(\omega-\Omega)
 \tilde E_{m\vecs K}(\Omega)\tilde\phi_{\vecs k'+\vecs K}(\omega'+\Omega),
 \label{calE}
\eea
where
\beq
 \tilde E_{m\vecs K}(\Omega)\equiv \int d^3xdt E_m(x)
e^{-i\vecs K\cdot \vecs x +i\Omega t}. \label{EFourier}
\eeq
Equation (\ref{calE}) should be compared with the Fourier transform of the
two-point correlation of the multiplicative noise, 
$\CD(x,x')\equiv \phi(x)\langle \xi_m(x)\xi_m(x')\rangle \phi(x')
=\phi(x)D_m(x-x')\phi(x')$, which reads
\bea
 \tilde\CD_{\vecs k, \vecs k'}(\omega,\omega')
 &\equiv& \int d^3xdt\int d^3x'dt'
  \CD(x,x')e^{-i\vecs k\cdot \vecs x +i\omega t}
 e^{-i\vecs k'\cdot \vecs x' +i\omega' t'} \non\\
 &=&\int \frac{d^3K}{(2\pi)^3}\frac{d\Omega}{2\pi}
 \tilde\phi_{\vecs k-\vecs K}(\omega-\Omega)
 \tilde D_{m\vecs K}(\Omega)\tilde\phi_{\vecs k'+\vecs K}(\omega'+\Omega).
 \label{calD}
\eea
Here $\tilde D_{m\vecs K}(\Omega)$ is defined in the same way as
(\ref{EFourier}). 

The multiplicative noise and dissipation under consideration are
generated by two graphs, $ L_3$ and $ L_8$, in the effective
action.  Since the relevant kernels $A_i(x-x')$ and $B_i(x-x')$
$(i=3,8)$, namely (\ref{33}), (\ref{37}), (\ref{39}), and (\ref{43}),
have the
same structure as in the case of three body bosonic interaction,
$ L_5$ or (\ref{35}) and (\ref{41}), which has been discussed in \S\ref{threebody}, we can easily
obtain the Fourier transform of the kernels. 

 Let us first 
consider the contribution of $ L_3$ to 
$\tilde C_{m \vecs K}(\Omega)=-i\Omega\tilde E_{m\vecs K}(\Omega)$ 
and $\tilde D_{m\vecs K}(\Omega)$.  From (\ref{Cboson}) and
(\ref{Dboson}) we find
\newpage
\bea
 \tilde{C}_{m\vecs K}(\Omega)
&=&-i\Omega\tilde E_{m\vecs K}(\Omega)
=\tilde A_{3\vecs K}(\Omega) \non\\
&=&-i\pi\frac{\lambda^2}{8}\int\frac{d^3q}{(2\pi)^3}
\frac{1}{\omega_q\omega_{K-q}} \non\\
&&~~~~~~~~\times\lnk
\lkk (1+n_q)(1+n_{K-q})-n_qn_{K-q}\rkk
\delta(\Omega-\omega_q-\omega_{K-q})\right. \non \\
&&~~~~~~~~~~
+\lkk (1+n_q)n_{K-q}-(1+n_{K-q})n_q\rkk
\delta(\Omega-\omega_q+\omega_{K-q})   \non \\
&&~~~~~~~~~~+\lkk n_{q}(1+n_{K-q})-(1+n_{q})n_{K-q}\rkk
\delta(\Omega+\omega_q-\omega_{K-q}) \non \\
&&~~~~~~~~~~\left.
+\lkk n_qn_{K-q}-(1+n_q)(1+n_{K-q})\rkk
\delta(\Omega+\omega_q+\omega_{K-q})\rnk,  \label{Cbosonm}
\eea
and
\bea
\tilde D_{m \vecs K}(\Omega)= \tilde B _{3\vecs K}(\Omega)&=&
\frac{\pi\lambda^2}{8}
\int\frac{d^3q}{(2\pi)^3}
\frac{1}{\Omega_q\omega_{K-q}} \non \\
&&\times\lnk
\lkk (1+n_q)(1+n_{K-q})+n_qn_{K-q}\rkk
\delta(\Omega-\omega_q-\omega_{K-q})\right. \non \\
&&
+\lkk (1+n_q)n_{K-q}+(1+n_{K-q})n_q\rkk
\delta(\Omega-\omega_q+\omega_{K-q})   \non \\
&&+\lkk n_{q}(1+n_{K-q})+(1+n_{q})n_{K-q}\rkk
\delta(\Omega+\omega_q-\omega_{K-q}) \non \\
&&\left.
+\lkk n_qn_{K-q}+(1+n_q)(1+n_{K-q})\rkk
\delta(\Omega+\omega_q+\omega_{K-q})\rnk, \label{Dbosonm}
\eea
where $\omega_q\equiv \sqrt{\veq^2+m_\phi^2}$,
$\omega_{K-q}\equiv\sqrt{(\veK-\veq)^2+m_\phi^2}$, 
$n_q\equiv n_B(\omega_q)$, and $n_{K-q}\equiv n_B(\omega_{K-q})$.
We can read off destruction terms $ R_D$ and creation terms $ R_C$
of $\phi$ as in \S \ref{threebody} with
$ R_C/ R_D=e^{-\beta\Omega}$.  Hence we obtain
\beq
  \frac{\tilde D_{m \vecs K}(\Omega)}{\tilde \Gamma_{m \vecs K}(\Omega)}   
 = \frac{2\Omega}{i}
\frac{\tilde D_{m \vecs K}(\Omega)}{\tilde C_{m \vecs K}(\Omega)}   
=2\Omega\frac{e^{\beta\Omega}+1}{e^{\beta\Omega}-1}.
\eeq
This is twice the corresponding results for
additive
noises, (\ref{DGamma}), (\ref{DGammab}), and (\ref{DGammas}).
But this discrepancy is compensated by an additional factor 2 in
(\ref{calE}), so we can see that the same relation holds between the
noise dispersion and the actual dissipation rate for the multiplicative case
as in the case of additive noises. Hence the generalized
 fluctuation-dissipation relation is satisfied in this case, too, to
 establish thermal equilibrium in the final state.

Note that 
contribution of $ L_8$ can also be manipulated by the replacement 
$\lambda^2\longrightarrow h^4$, $\omega_q\longrightarrow 
\sqrt{\veq^2+m_\chi^2}$, and 
$\omega_{K-q}\longrightarrow\sqrt{(\veK-\veq)^2+m_\chi^2}$.

\subsection{Dissipation rate of zero-mode oscillation}

So far we have shown that multiplicative noise and dissipation also
satisfy the desired fluctuation-dissipation relation generically, but we
cannot obtain the magnitude of the dissipation rate without knowing
the Fourier transform of the scalar field itself.  
Here we consider a specific field evolution and calculate the
dissipation rate.  To do this we consider the case only zero-mode
oscillation with a fixed angular frequency $M$ is present, namely,
we adopt the harmonic expansion as in (\ref{harmonic}),
\beq
 \phi(\vex,t-\tau)=
 \phi(t-\tau)=\phi(t)\cos M\tau -\frac{\dot\phi(t)}{M}\sin M\tau,
 \label{harmonic2}
\eeq
which is the case we are most interested in.   For this approximate 
solution to be
valid we assume that $\lambda\phi^3/3!$ is smaller than the mass term
$M^2\phi$ in the equation of motion (\ref{multieq2}).
Then the last term of the left-hand-side of Eq.\ (\ref{multieq2}), which
represents the dissipative effects, reads
\bea
&&\int_{-\infty}^{t}dt'\int d^3x'
             2\phi(x) E_{m}(x-x')\phi(x')=\int_0^\infty d\tau 
\int \frac{d\omega}{2\pi} 2\tilde E_{m\vecs 0}(\omega) 
\phi(t)\phi(t-\tau)\dot\phi(t-\tau) \non\\
&=&\int_0^\infty d\tau \int \frac{d\omega}{2\pi} 
2\tilde E_{m\vecs 0}(\omega)e^{-i\omega\tau}\phi(t)
\lkk \phi(t)\dot\phi(t)\cos 2M\tau - \frac{1}{2}
\lmk \frac{\dot\phi^2(t)}{M}-M\phi^2(t)\rmk\sin 2M\tau\rkk \non\\
&\equiv& \Lambda_1\phi^2(t)\dot\phi(t)
+\Lambda_2 \lmk\frac{\dot\phi^2(t)}{M}-M\phi^2(t)\rmk\phi(t)
\eea
where
\bea
\Lambda_1&\equiv&\int_0^\infty d\tau \int \frac{d\omega}{2\pi} 
2\tilde E_{m\vecs 0}(\omega)e^{-i\omega\tau}\cos 2M\tau 
= \frac{1}{2}\lkk \tilde E_{m\vecs 0}(2M)+\tilde E_{m\vecs 0}(-2M)\rkk
 \\
&=&\frac{\lambda^2}{64\pi
M}\lkk 1-\lmk\frac{m_\phi}{M}\rmk^2\rkk^{1/2}\lkk 1+2n_B(M)\rkk, \non\\
\Lambda_2&\equiv&-\int_0^\infty d\tau \int \frac{d\omega}{2\pi} 
\tilde E_{m\vecs 0}(\omega)e^{-i\omega\tau}\sin 2M\tau .
\eea
Multiplying the effective equation of motion, 
\beq
\ddot\phi(t)+M^2\phi(t)+\Lambda_1\phi^2(t)\dot\phi(t)
+\Lambda_2 \lmk\frac{\dot\phi^2(t)}{M}-M\phi^2(t)\rmk\phi(t)=0,
\eeq
by $\dot\phi(t)$, we find
\beq
 \frac{d\rho_{\phi}(t)}{dt}\equiv
 \frac{d~}{dt}\lmk \frac{1}{2}\dot\phi^2(t)+\frac{1}{2}M^2\phi^2(t)\rmk
 =-\Lambda_1\phi^2(t)\dot\phi^2(t)-\Lambda_2\lmk \frac{\dot\phi^2(t)}{M}
-M\phi^2(t)\rmk\phi(t)\dot\phi(t). \label{energy}
\eeq
So far $\phi(t)$ is a value at an arbitrary time $t$ around which 
the harmonic expansion (\ref{harmonic2}) is performed.  The
right-hand-side of the above equation severely depends on the phase of
the scalar field at the time $t$.  Hence we take an average over the
phase of the oscillation to obtain its typical magnitude.  As a result we find
the second term vanishes and (\ref{energy}) reads
\beq
 \frac{d\rho_{\phi}(t)}{dt}=-\frac{1}{2}\Lambda_1\phi^2(t)\rho_{\phi}(t),
\eeq
where $\phi^2(t)$ should now be interpreted as a
 mean square amplitude around the time $t$
rather than its instantaneous  value then.

Thus the dissipation rate is given by
\beq
  \tilde\Gamma_0=\frac{\lambda^2\phi^2(t)}{128\pi
M}\lkk 1-\lmk\frac{m_\phi}{M}\rmk^2\rkk^{1/2}\lkk 1+2n_B(M)\rkk.
\label{multidispphi} 
\eeq

Similarly, the dissipation rate associated with $h^2\phi^2\chi^2/4$
interaction represented by the graph $ L_8$ reads
\beq
 \tilde\Gamma_0=\frac{h^4\phi^2(t)}{128\pi
M}\lkk 1-\lmk\frac{m_\chi}{M}\rmk^2\rkk^{1/2}\lkk 1+2n_B(M)\rkk\equiv
\Gamma_M(T). 
\label{multidispchi}
\eeq
Although the interaction $h^2\phi^2\chi^2/4$ represents  creation of a
pair of $\chi$ from pair annihilation of $\phi$ formally, the coherent
nature of field oscillation makes it possible to interpret
 the above dissipation rate just as a decay of
$\phi$ particle with oscillating frequency $2M$ through three-body
bosonic interaction $\CM\phi\chi^2$ with the coupling strength $\CM=h^2\phi/2$.
They are valid for 
 $\lambda\phi^2\lesssim M^2$ and $h|\phi| \lesssim M$.

\section{Application to the late reheating phase of the inflationary universe}
\label{inflation}
So far we have studied relaxation of an oscillating scalar field
through various interaction channels and obtained the dissipation rate
or the relaxation time scale to thermal equilibrium for each case.  We now
apply our results to two cosmological problems, one the reheating after
inflation \cite{Sato:1980yn,lindebook}
and the other evaporation of oscillating quasiflat direction
in supersymmetric theory in relation with Affleck-Dine baryogenesis 
\cite{Affleck:1984fy}.  In this section we consider the former problem
and the latter will be discussed in the next section.

\subsection{Brief review of previous results}
 
First we list several useful formulae of reheating after inflation which
has been studied extensively in the literatures 
\cite{Albrecht:1982mp,Dolgov:1982th,Traschen:1990sw,Kofman:1994rk}.  
Slow-roll inflation is terminated
as it is followed by coherent scalar field oscillation, whose energy
density is released to that of radiation subsequently.
Two mechanisms are known to reheat the universe.  One, which can be very
efficient, is parametric resonance dubbed {\it
preheating} \cite{Traschen:1990sw,Kofman:1994rk,Kolb:1996jt}.   
As its name tells, however, it is effective only in the
early stage of reheating when the inflaton scalar field is oscillating with a
sufficiently large amplitude and only when it is coupled to other scalar
fields.   The other is perturbative decay of the inflaton field which
terminates reheating process.  One loop calculation shows that the
dissipation rate is equal to the decay rate of the inflaton $\phi$ 
 \cite{Dolgov:1982th,Dolgov:1998wz}.

For a constant value of the decay rate $\Gamma_\phi$, the energy density
of the oscillating inflaton, $\rhophi(t)$, and that of radiation,
$\rhorad(t)$, satisfy the following transfer equations.
\bea
  \frac{d\rhophi(t)}{dt}&=&-(3H+\Gamma_\phi)\rhophi(t), \label{rhophi}\\
  \frac{d\rhorad(t)}{dt}&=&-4H\rhorad(t)+\Gamma_\phi\rhophi(t), \label{rhorad}
\eea
which are valid when parametric resonance is unimportant.  We are also
assuming that the scalar field oscillation is driven by its mass term
and higher order interactions are negligible,
namely, 
\beq
  \rhophi(t)=\frac{1}{2}\dot\phi^2(t)+\frac{1}{2}M^2\phi^2(t). 
\label{rhophidef}
\eeq
The solution of (\ref{rhophi}) and (\ref{rhorad}) are then given by
\bea
  \rhophi(t)&=&\rhophi(t_i)\lmk\frac{a(t)}{a(t_i)}\rmk^{-3}\!\!\!\!
e^{-\Gamma_\phi(t-t_i)}, \label{rhophisol}\\
  \rhorad(t)&=&\rhorad(t_i)\lmk\frac{a(t)}{a(t_i)}\rmk^{-4}
  +\Gamma_\phi\int_{t_i}^{t}\lmk\frac{a(t)}{a(\tau)}\rmk^{-4}\!\!\!\!
  \rhophi(\tau)d\tau. \label{rhoradsol}
\eea
Here $t_i$ is the time when parametric resonance becomes no longer effective
or the epoch when the inflaton starts coherent oscillation after
inflation, whichever comes later.  In the latter case we take
$\rhorad(t_i)=0$ of course.  In the above system the scalar field decays
around   $t\simeq \Gamma_\phi^{-1}$ and reheating is completed.  
For definiteness we define the reheating epoch by the time when the
Hubble parameter $H$ becomes equal to $\Gamma_\phi$, so that the reheat
temperature, $T_R$, reads
\beq
  T_R=\lmk\frac{90}{\pi^2g_\ast}\rmk^{1/4}\sqrt{\Gamma_\phi\mg}
  =0.46\tildegs^{-1/4}\sqrt{\Gamma_\phi\mg}, \label{reheat}
\eeq
where $\mg=2.4\times 10^{18}$GeV is the reduced Planck mass, and $g_\ast$
is the effective number of the relativistic degrees of freedom with
$\tildegs\equiv g_\ast/200$. 

Note, however, that this is not the maximum temperature after inflation
but that when entropy production from the inflaton
is practically terminated.  Even when
preheating is inoperative, the maximum temperature can be much higher
than $T_R$ as we can write (\ref{rhoradsol}) as
\beq
  \rhorad(t)=\frac{3}{5}\Gamma_\phi t \lmk\frac{a(t)}{a(t_i)}\rmk^{-3}
\!\!\!\!
\rhophi(t_i)\cong \frac{6}{5}\Gamma_\phi H\mg^2,  \label{rhoradgamma}
\eeq
for $t_i \ll t \ll \Gamma_\phi^{-1}$ with $\rhorad(t_i)=0$.  That is, if
the decay product of inflaton is rapidly thermalized, the cosmic
temperature in the field oscillation regime without preheating is given
by
\beq
 T\cong\lmk\frac{36}{\pi^2g_\ast}\Gamma_\phi
 H\mg^2\rmk^{1/4}. \label{temperature}
\eeq
Note that this expression is valid well until the reheating time
$H=\Gamma_\phi$ when (\ref{temperature}) agrees with (\ref{reheat})
with an error of 26\%.  From (\ref{reheat}) and (\ref{temperature})
we obtain a formula 
\beq
  T\cong 0.54\tildegs^{-1/8}(T_R^2H\mg)^{1/4}\simeq  (T_R^2H\mg)^{1/4},
  \label{temperature2}
\eeq
which will be useful later.

The above is the case with a constant $\Gamma_\phi$.
We now consider the cases dissipation rate of the inflaton is given by
our new results with possible temperature dependence.
As in (\ref{lagrangian}) we take the interaction Lagrangian as
\beq
-\CL_{\rm int} = \frac12\,M^2\phi^2 +
     \frac{1}{4!}\lambda\,\phi^4
     +\frac12\,m_{\chi}^2\chi^2 +
    \CM \phi\chi^2 +
     \frac{1}{4}\,h^2 \chi^2 \phi^2
       + m_{\psi} \bar\psi \psi
     + f\phi\bar \psi\psi \:,
\eeq
where we assume $m_\chi$ and $m_\psi$ are much smaller than the inflaton mass,
$M$, and neglect them in the subsequent discussion.
As before, $M^2$ includes both intrinsic mass $m_\phi^2$ and
high-temperature corrections of order of $\sim h^2T^2$ and/or $\sim
f^2T^2$.  These thermal masses are present if the oscillating masses of
$\chi$ and $\psi$ are smaller than the temperature, namely, $h|\phi|< T$
and $f|\phi| <T$, respectively.  We therefore find $h^2T^2\phi^2 < T^4$
and $f^2T^2\phi^2 < T^4$.  Since the energy density of oscillating
inflaton remains larger than that of radiation up to the reheating time,
these inequalities mean thermal masses are smaller than $m_\phi$, so
$M=m_\phi$ in this regime.

It should be understood that the above form of the interaction
Lagrangian is a result of expansion around the potential minimum which
we have set to $\phi=0$ after an appropriate shift, if necessary.  
Hence the values of
the parameters may not be fixed by the amplitude and spectrum of density
fluctuations straightforwardly.  In particular, if inflation occurred
more than once, the parameters of the last inflation may entirely
be free from large-scale observations.  
If, on the other hand,
 it describes the original potential as it is and if 
chaotic inflation \cite{Linde:gd}
was driven by $\phi$, we find $M\simeq 10^{13}$GeV and
$\lambda \lesssim 10^{-13}$ \cite{Salopek:1992zg}.  Then in order that
radiative corrections do not disturb the potential we require $h\lesssim
10^{-3}$, $f \lesssim 10^{-3}$ and $\CM \lesssim 10^{13}$GeV.

Below we consider the effect of each interaction term separately.

\subsection{Reheating through Yukawa coupling} 

First we consider the case the inflaton is coupled only with fermions
$\psi$ and $\overline\psi$
through Yukawa coupling.  In this case preheating due to parametric
resonance is unimportant and the dissipation rate is given from 
(\ref{fermiondissipate}) as
\beq
  \Gamma_F(T)=\frac{f^2}{8\pi}M\lkk 1-
2n_F\lmk\frac{M}{2}\rmk\rkk, \label{fermiondissipate2}
\eeq
in the perturbative regime $M\gtrsim f\phi$.  If the reheat temperature
turns out to be much lower than $M$ the dissipation rate agrees with the
conventional calculation which gives one particle decay rate of $\phi$,
\beq
  \Gamma_{F,\rm conv}=\frac{f^2}{8\pi}M,
\eeq
which gives
\beq
  T_{R,\rm conv}=\lmk\frac{90}{\pi^2 g_\ast}\rmk^{1/4}
\lmk\frac{f^2M\mg}{8\pi}\rmk^{1/2} =4.5\times 10^{11}\tildegs^{-1/4}
\lmk\frac{f}{10^{-3}}\rmk\lmk\frac{M}{10^{13}\GeV}\rmk^{1/2}\GeV.
  \label{convreheatf}
\eeq
On the other hand, if $M$ is so small that
the reheating is completed in a high temperature
regime $T \gg M$, we find from (\ref{fermiondissipate2}) that 
\beq
  \Gamma_{F,\rm high}(T)=\frac{f^2}{32\pi}\frac{M^2}{T}.
\eeq
Inserting it in (\ref{reheat})
 the reheat temperature is approximately given by
\beq
  T_{R,\rm high}\cong \lmk\frac{90}{\pi^2 g_\ast}\rmk^{1/6}
\lmk\frac{f^2M^2\mg}{32\pi}\rmk^{1/3}.  \label{newreheatf}
\eeq
This formula applies when $T_{R,\rm high} \gg M$, or
\beq
  M \ll \lmk\frac{90}{\pi^2 g_\ast}\rmk^{1/2}\frac{f^2\mg}{32\pi}
 =5\times 10^9\tildegs^{-1/2}\lmk\frac{f}{10^{-3}}\rmk^2 \GeV. 
\label{Mcond}
\eeq
In this case relaxation of the inflaton is delayed due to Pauli blocking.
But the discrepancy between the new result (\ref{newreheatf}) and the
conventional one (\ref{convreheatf}) is rather modest,
\beq
  \frac{T_{R,\rm high}}{T_{R,\rm conv}}=0.5\tildegs^{-1/2}
 \lmk\frac{f}{10^{-3}}\rmk^{-1/3}\lmk\frac{M}{5\times
 10^9\GeV}\rmk^{1/6}, 
\eeq
with a weak dependence on the model parameters $(Mf^{-2})^{1/6}$.

Finally we confirm consistency of our analysis.
The condition  $f\phi<M$ is satisfied at the time of
reheating for
\beq
 M > 4\times 10^6\tildegs^{1/4}\lmk\frac{f}{10^{-3}}\rmk^{7/2}\GeV.
 \label{cf1}
\eeq
On the other hand, the condition that thermal mass of $\phi$ generated
by Yukawa coupling, $fT$, is smaller than $M$ reads
\beq
  M > 5\times 10^6\tildegs^{-1/2}\lmk\frac{f}{10^{-3}}\rmk^{3}\GeV.
\label{cf2}
\eeq
We see that (\ref{Mcond}), (\ref{cf1}), and (\ref{cf2})
can easily be satisfied simultaneously. 

\subsection{Reheating through three body bosonic interaction}

Next we consider the three body bosonic interaction $\CM \phi\chi^2$,
which induces a dissipation rate (\ref{chidecay})
\beq
  \Gamma_B(T)=\frac{\CM^2}{8\pi M}
\lkk 1+2n_B\lmk\frac{M}{2}\rmk \rkk,
\eeq
In the high temperature limit   the dissipation rate is enhanced as
\beq
 \Gamma_{B, \rm high}(T) = \frac{\CM^2T}{2\pi M^2}.
 \label{bosonhigh}
\eeq
This expression applies when $T \gg M$ and
 $M^2 > \CM\phi$.  The latter requirement is the same as the condition
 broad resonance
is no longer effective.   
We are interested in the case reheating is completed in this high
temperature regime.  Using the formula
\beq
  T_{R,\rm high}\cong \lmk\frac{90}{\pi^2
  g_\ast}\rmk^{1/4}\sqrt{\Gamma_{B, \rm high}(T_{R,\rm high})\mg},
\label{reheatformula}
\eeq
we would obtain the reheat temperature
\beq
  T_{R,\rm high}=\lmk\frac{90}{\pi^2
  g_\ast}\rmk^{1/2}\frac{\CM^2\mg}{2\pi M^2}. \label{bosonreheathigh}
\eeq
Consistency $T_{R,\rm high}\gg M$ would then read
\beq
  M \ll \lmk\frac{90}{\pi^2
  g_\ast}\rmk^{1/6}\lmk\frac{\CM^2\mg}{2\pi}\rmk^{1/3}=2.0\times 10^{12}
 \lmk\frac{\CM}{10^{10}\GeV}\rmk^{2/3} \GeV. \label{bosonconsistency}
\eeq
The other condition for (\ref{bosonhigh}) to apply, namely $\CM\phi <
M^2$, requires the radiation energy with temperature (\ref{bosonreheathigh})
should be smaller than $M^6/\CM^2=\rhophi(\phi=M^2/\CM)$, which reads
\beq
 M>1.3\times 10^{12}\tildegs^{-1/14}
\lmk\frac{\CM}{10^{10}\GeV}\rmk^{5/7}\GeV.
 \label{bosonconsistency2}
\eeq
Clearly, (\ref{bosonconsistency}) and (\ref{bosonconsistency2}) are
hardly compatible with each other.

This means that if reheating is governed by the high-temperature
dissipation rate (\ref{bosonhigh}) the reheating process occurs shortly
after the field amplitude gets smaller than $M^2/\CM$ when
(\ref{bosonhigh}) becomes applicable.
\if
For the moment let us pretend 
that preheating is negligible
 and all the radiation comes from dissipation (\ref{bosonhigh}) to yield
radiation density and the temperature for $\phi < M^2/\CM$ 
\beq
  \rhorad\simeq \Gamma_{\phi, \rm high}(T)H\mg^2,
~~T\simeq \lmk\frac{30}{\pi^2  g_\ast}\rmk^{1/3} 
\lmk\frac{\CM^2\mg^2 H}{2\pi M^2}\rmk^{1/3},  \label{faketemp}
\eeq
where we have used (\ref{rhoradgamma}), which is not strictly valid when 
the dissipation rate depends on background temperature but still gives
reasonably correct order of magnitude.  
Inserting (\ref{bosonreheathigh}) to (\ref{bosonhigh}) and comparing the
resultant dissipation rate with the Hubble parameter at the epoch $\phi
\lesssim M^2/\CM$ when (\ref{bosonhigh}) becomes applicable, we find
that $\Gamma_{\phi, \rm high}(T) > H\simeq \frac{M^3}{\sqrt{3}\mg \CM}$ holds
if
\beq
  M< \lmk\frac{90}{\pi^2
  g_\ast}\rmk^{1/12}\lmk\frac{\CM^2\mg}{(2\pi)^{3/4}}\rmk^{1/3}
 =2.0\times 10^{12}
 \lmk\frac{\CM}{10^{10}\GeV}\rmk \GeV.
\eeq

This inequality is automatically 
satisfied when (\ref{bosonconsistency}) is satisfied.

If we incorporate the contribution of radiation produced by broad
parametric resonance, we would obtain higher temperature than
(\ref{faketemp}) and higher dissipation rate but the Hubble parameter at
the time $\phi=M^2/\CM$ would not increase too much.  As a result the
inequality $\Gamma_{\phi, \rm high}(T) > H$ will more easily be
satisfied at the epoch $\phi=M^2/\CM$.

Thus we may conclude that in the 
situation the temperature-enhanced dissipation rate
(\ref{bosonhigh}) applies, 
 the scalar field is dissipated within the Hubble time
as soon as (\ref{bosonhigh}) becomes applicable as the field amplitude
becomes adequately smaller than $M^2/\CM$.
\fi
Then the
use of the formula (\ref{reheatformula}) is inappropriate and we should
use 
\bea
 T_{R,\rm high}&\simeq& \lmk\frac{90}{\pi^2
  g_\ast}\rmk^{1/4}\sqrt{H_c\mg}=
\lmk\frac{30\xi^2}{\pi^2  g_\ast}\rmk^{1/4} \frac{M^{3/2}}{\CM^{1/2}} \non\\
&=&3.5\times 10^{12}\tildegs^{-1/4}\xi^{1/2}
\lmk\frac{M}{10^{12}\GeV}\rmk^{3/2}
\lmk\frac{\CM}{10^{10}\GeV}\rmk^{-1/2}\GeV, \label{newbosonreheat}
\eea
where $H_c$ denotes the Hubble parameter when $\phi$ becomes as small as
$M^2/\CM$, namely
\beq
  H_c = \frac{\xi M^3}{\sqrt{3}\mg\CM}.
\eeq
Here $\xi \geq 1$ is a parameter which represents
contribution of residual radiation energy density $\rhorad$ created by the
parametric resonance. It is defined by $\rhorad = (\xi^2-1)\rhophi$ and
would reduce to unity if preheating was totally
negligible. 
Now the consistency condition $T_{R,\rm high}\gg M$ reads
\beq
  M\gg \lmk\frac{\pi^2  g_\ast}{30}\rmk^{1/2}\xi^{-1}\CM
  =8.1\tildegs^{1/2}\xi^{-1}\CM. 
\eeq
Let us confirm the dissipation rate (\ref{bosonhigh}) at the
temperature (\ref{newbosonreheat}) is larger than $H_c$,
which yields
\beq
  M < 1.3\times 10^{12}\tildegs^{-1/14}\xi^{-1/7}
\lmk\frac{\CM}{10^{10}\GeV}\rmk^{5/7}\GeV,
\eeq
which is consistent with (\ref{bosonconsistency2}), because these two
inequalities have been derived from the opposite conditions.

On the other hand, if the conventional dissipation rate
$\Gamma_{\phi,\rm conv}=\CM^2/(8\pi M)$ was larger than $H_c$,
or
\beq
  M< 6.4\times 10^{11}\xi^{-1/4}\lmk\frac{\CM}{10^{10}\GeV}\rmk^{3/4}\GeV,
\eeq
 the conventional reheating process would also proceed as rapidly as to
 give the same reheat temperature.  Hence the effects of
the high-temperature
 enhancement of the dissipation rate is prominent only when the inequality
\beq
  6.4\times 10^{11}\xi^{-1/4}\lmk\frac{\CM}{10^{10}\GeV}\rmk^{3/4}\GeV < M <
1.3\times 10^{12}\tildegs^{-1/14}\xi^{-1/7}
\lmk\frac{\CM}{10^{10}\GeV}\rmk^{5/7}\GeV,
\eeq
is satisfied.
As a result the ratio of the new reheat temperature
(\ref{newbosonreheat})
to the conventional estimate,
\beq
  T_{R,\rm conv}=\lmk\frac{90}{\pi^2
  g_\ast}\rmk^{1/4}\lmk\frac{\CM^2\mg}{8\pi M}\rmk^{1/2}
 =1.4\times 10^{12}\tildegs^{-1/4}\lmk\frac{\CM}{10^{10}\GeV}\rmk
 \lmk\frac{M}{10^{12}\GeV}\rmk^{-1/2}\GeV,  \label{convreheatb}
\eeq
is at most
\beq
  \frac{T_{R,\rm high}}{T_{R,\rm conv}}=\frac{\sqrt{8\pi\xi}}{3^{1/4}}
 \frac{M^2}{\CM^{3/2}\mg^{1/2}}<4.2\tildegs^{3/14}\xi^{-1/7}
 \lmk\frac{\CM}{10^{10}\GeV}\rmk^{-1/14}.
\eeq

\subsection{Reheating through setting-sun diagrams}

Finally we consider reheating through dissipation due to the setting-sun
diagrams, in particular, arising from the interaction
$h^2\phi^2\chi^2/4$ corresponding to the diagram $ L_7$.  The
dissipation rate which applies at high temperature $T \gg M=m_\phi$ and low
field amplitude $\phi \lesssim M/h$ after the broad resonance regime 
 is (\ref{sunsetgammam}) or the second line of (\ref{settingsungamma}),
\beq
  \Gamma_{S}(T) = \frac{3h^4T^2}{256\pi M}, 
 \label{sunsetgamma2}
\eeq
which has the same temperature dependence as the Hubble parameter in the
radiation dominated universe.  Hence in order to reheat the universe
completely due to this dissipation term we must have
\beq
  \Gamma_{S}(T) = \frac{3h^4T^2}{256\pi M} > 
\lmk\frac{\pi^2 g_\ast}{90}\rmk^{1/2}\frac{T^2}{\mg},
\eeq
namely,
\beq
  M < 1.9\times 10^{3}\tildegs^{-1/2}\lmk\frac{h}{10^{-3}}\rmk^4\GeV.
\label{masscond}
\eeq

Let us first pretend  that preheating is negligible
 and all the radiation comes from dissipation (\ref{sunsetgamma2}). Then
radiation density and the temperature for $\phi < M/h$ are given by
\beq
  \rhorad\simeq \Gamma_{S}(T)H\mg^2,
~~T\simeq \lmk\frac{30}{\pi^2  g_\ast}\rmk^{1/2} 
\lmk\frac{h^4\mg^2 H}{64\pi M}\rmk^{1/2}.  \label{faketempss}
\eeq
Here we have used (\ref{rhoradgamma}), which is not strictly valid when 
the dissipation rate depends on background temperature but still gives
reasonably correct order of magnitude.  
Inserting (\ref{faketempss}) to (\ref{sunsetgamma2})
we find
\beq
  \Gamma_{S}(T)\simeq \frac{30}{\pi^2  g_\ast}
\lmk\frac{h^2\mg}{64\pi M}\rmk^2H > \frac{H}{3},
\eeq
when the inequality (\ref{masscond}) is satisfied.  
Thus we find that $\Gamma_{S}(T)$ is already close to $H$ even
 if we take into account only the radiation produced by perturbative
 processes governed by (\ref{sunsetgamma2}), and in this case
it can be larger than
 $H$ when
\beq
  M < 1.1\times 10^3\tildegs^{-1/2}\lmk\frac{h}{10^{-3}}\rmk^4\GeV,
\eeq
to reheat the universe soon after the epoch $\phi \lesssim M/h$.

If we include the effect of preheating 
 the cosmic temperature could be 
higher than (\ref{faketempss}).  Then $\Gamma_{S}(T)$ could be
larger than $H$ at the epoch $\phi \lesssim M/h$ under (\ref{masscond}).
For this to be the case, preheating is only required to create
twice or more radiation than perturbative processes during
 broad resonance regime.

Finally we examine the consistency of our analysis, $hT_R < M \ll T_R$,
where the former is the condition that thermal mass of $\phi$
generated through $h^2\phi^2\chi^2/4$ interaction remains smaller than
$M$.  Denoting the residual radiation energy density due to preheating
by $\rhorad=(\xi^2-1)M^2\phi^2$ as in the previous subsection, the reheat
temperature $T_{R,S}$ reads,
\beq
  T_{R,S}\simeq \lmk\frac{30}{\pi^2
  g_\ast}\rmk^{1/4}\lmk\frac{\xi}{h}\rmk^{1/2}M=11\tildegs^{-1/4}
 \lmk\frac{h}{10^{-3}}\rmk^{-1/2}\xi^{1/2}M, 
\eeq
because the total energy density, $\rho_{\rm tot}=\xi^2 M^2\phi^2$, is
efficiently converted to radiation at $\phi\simeq M/h$ in this scenario.
Thus the desired condition is easily satisfied.

Due to the strong dependence of the dissipation rate on the coupling
constant $h^4$, the above processes are operative only for inflation
with a small mass scale $M\sim 10^3$GeV for small coupling
$h\sim 10^{-3}$.  It is interesting to note, however, that in this case
the scalar
field can dissipate its energy to get  thermalized even in the absence
of interactions that lead $\phi$ to decay, such as $f\phi\psi\overline\psi$
or $\CM \phi\chi^2$.

\section{Evaporation rate of oscillating flat direction}
\label{flatdirection}
\subsection{Behavior of flat direction after inflation}

In generic supersymmetric theories there are a number of directions in
scalar field configuration space along which the potential vanishes
except for a soft supersymmetry-breaking mass term.  Such a flat direction
field may acquire a large expectation value  of order of $\mg$, beyond
which the potential blows up exponentially in minimal supergravity, by 
accumulating
quantum fluctuations during inflation and they start coherent field
oscillation only after the Hubble parameter has decreased to the soft
mass of order of $\sim$TeV or so.  Then the large-amplitude
oscillation can easily violate baryon and lepton number conservation to
generate baryon-to-entropy ratio 
  up to $\order (1)$.  This is the
original picture of the Affleck-Dine baryogenesis \cite{Affleck:1984fy}
supplemented by inflationary cosmology \cite{Linde:gh}. 

 Later  the
effect of finite-density supersymmetry breaking, especially, the
Hubble-induced mass term and the
importance of the nonrenormalizable terms in the superpotential 
were investigated by Dine, Randall and Thomas \cite{Dine:1995kz}. They
included the following nonrenormalizable term in the superpotential $W$.
\beq
  W \supset \frac{\lambda_n}{n\mst^{n-3}}\varphi^n,
\eeq
where $\varphi$  denotes a flat direction field, $n$ is an integer
larger than 3,
$\lambda_n$ is a
constant of order of unity, $\mst$ is some large 
cut-off  scale
such as the GUT or Planck scale.
   Together with the Hubble-induced mass term, the scalar
potential  reads
\beq
  V(\varphi)\simeq  m_\varphi^2|\varphi|^2-cH^2|\varphi|^2
 +\lkk \frac{(A\mgv+aH)\lambda_n\varphi^n}{n\mst^{n-3}} +{\rm H.C.}\rkk 
 +|\lambda_n|^2\frac{|\varphi|^{2n-2}}{\mst^{2n-6}}, \label{pote}
\eeq
where $A$, $a$, and $c$ are dimensionless quantities of order of unity,
and $\mgv\sim 1$TeV is the gravitino mass which we expect is of the same
order of  the soft mass 
$m_\varphi$.  It is important to have $c$ negative.
Then the instantaneous minimum of $\varphi$ is located at
\beq
 |\varphi|
 \simeq (H\mst^{n-3})^{1/(n-2)},  \label{varphiexpect}
\eeq
when $H \gg m_{3/2}\sim m_\varphi$.  

In the scenario of Dine, Randall, and
Thomas \cite{Dine:1995kz}, the scalar field starts oscillation with the
angular frequency $m_\varphi$ as $H$ becomes less than $m_\varphi\sim
m_{3/2}$, and baryon number is generated.  At this stage, however, there
was a fear that the scalar condensate might evaporate before sufficient
oscillation was achieved, because they postulated that the Affleck-Dine
field would evaporate due to the scattering by thermal particles produced
during the inflaton oscillation regime with the temperature
(\ref{temperature}).  As shown in \cite{Anisimov:2000wx} the scattering
crosssection of zero-mode particle with mass $m_\varphi\sim m_{3/2}$ by a
thermal particle such as a fermion with Yukawa coupling $f$
with energy and momentum $\sim T$ is of order of
\beq
  \sigma \sim \frac{f^2\alpha}{m_{3/2}T},  \label{crosssection}
\eeq
where $\alpha=g^2/(4\pi)$ is a gauge coupling strength.  Multiplying the number
density of thermal particle $n\sim T^3,$ the ratio of the scattering rate
$\Gamma_\varphi$ to $H$ reads
\beq
  \frac{\Gamma_\varphi}{H}\sim \frac{f^2\alpha T^2}{m_{3/2}H}
  \gtrsim \frac{f^2\alpha\mg}{m_{3/2}},
\eeq
where we have  neglected numerical factors, and the inequality is
saturated in the radiation dominated regime with $H\sim T^2/\mg$.  
Apparently this quantity is much larger than unity for reasonable values
of $m_{3/2}$, $f$, and $\alpha$. 

If the flat direction interacts with thermal particle as above, however,
its
potential acquires finite-temperature corrections such as a thermal mass
term $\sim fT$ at the same time.
  As a result the flat direction may start coherent
oscillation much earlier than previously assumed
\cite{Allahverdi:2000zd}.  Then the above estimate of the evaporation
rate does not apply, and the authors of \cite{Allahverdi:2000zd} used the
scattering rate with thermal particles, $\Gamma_\varphi\sim f^4T$ or 
$\Gamma_\varphi\sim g^4T$, for the evaporation rate of the flat
direction, 
where $g~( >f)$ is the gauge coupling. The 
former formula applies when $f|\varphi|< T<g|\varphi|$ and the latter
for
$g|\varphi|<T$.

These crude estimate has been refined by Anisimov and Dine
\cite{Anisimov:2000wx}.  They observed the center-of-mass energy between
zero mode condensate with mass $fT$ and a massless thermal particle 
is of order $f^{1/2}T$ whose square should replace the denominator of
(\ref{crosssection}). As a result they find 
\beq
  \Gamma_\varphi \sim f\alpha T. \label{scat}
\eeq

Whichever type of masses are used, in all the above estimates of the
evaporation rate of the oscillating flat directions, it was 
analyzed  with a picture of particle-particle scattering.
However, since the zero-mode  field oscillation occupies the
entire space homogeneously, it would be more appropriate to regard it as
a coherent condensate rather than particles.  
Hence  we should use the
formalism developed in the present paper instead.   

\subsection{Dissipation rates of  oscillating flat direction
 with a thermal mass}

Although flat direction fields are complex scalar fields, if the main
driving force of their oscillation is their mass term, we can
approximately regard them as a pair of independent real scalar fields
and use our results based on finite-temperature nonequilibrium field
theory to calculate the evaporation rate.  If, on the other hand,
$\varphi$ had a large initial value $\varphi\sim \mg$ and the condensate
acquired huge baryon number density, its evaporation would be delayed
because chemical potential of bosons cannot exceed their mass
\cite{Dolgov:sy,Dolgov:2002vf}. 
 We assume that initial value of $\varphi$ is
regulated to a sufficiently small value (\ref{varphiexpect}) due to the
nonrenormalizable terms in the potential (\ref{pote}) and consider the
situation the flat directions dissipate their energy through the relevant
dissipation rate we have obtained in \S III and \S IV.

These fields can possess all
types of interactions discussed so far, namely, Yukawa coupling 
$f\varphi\bar\psi \psi$, three-body scalar interaction $\CM
\varphi\chi^2$ and biquadratic interaction  
$h^2|\varphi|^2|\chi|^2$.  Here typical value of $\CM$ is $f\mu$ 
with $\mu$ being the energy scale of the standard model that emerges as
 the coefficient of $H_uH_d$ term in the superpotential of
the minimal supersymmetric standard model, while we expect several types
of biquadratic interactions with Yukawa coupling strength $h=f$ and 
gauge coupling strength $h=g > f$.  If the cosmic temperature is higher
than $g|\varphi|$ we expect $\varphi$ has a thermal mass of $\sim gT$
and it drives coherent oscillation when $H < gT$.  For $f|\varphi| < T <
g|\varphi|$ the flat direction has a thermal mass $\sim fT$ and it can
also drive oscillation when $H < fT$. 
Here we first write down the dissipation rates from various
interactions for each case and then consider which rates are applicable
in the next subsection.
  
First we consider the case 
 $T>g|\varphi|$ and $H < gT$ so that $\varphi$ is oscillating with the
 angular frequency $M=gT < T$.  Using 
 (\ref{fermiondissipate}), (\ref{chidecay}),
(\ref{settingsungamma}), and (\ref{multidispchi}), we can list the
 rate of each dissipation channel together with the range of its applicability.
\bea
 \Gamma_F &\cong&  \displaystyle{\frac{f^2M^2}{32\pi
 T}=\frac{f^2g^2}{32\pi}T},
 \qquad\qquad\qquad~~
 {\rm for~} f|\varphi| < gT,
  \label{gF}\\[4mm]
 \Gamma_B &\cong& \displaystyle{\frac{(f\mu)^2 T}{2\pi M^2}
=\frac{f^2\mu^2}{2\pi g^2 T}},
 \qquad\qquad\qquad {\rm for~} f\mu|\varphi| < g^2T^2, 
\eea
\bea
 \Gamma_S &\cong& \lnk\begin{array}{ll}
\displaystyle{\frac{g^4T^2}{64\pi M} = \frac{g^3}{64\pi}T},
 \qquad\qquad\qquad & \mbox{ for~} |\varphi| \lesssim T, \\[5mm]
\displaystyle{\frac{g^4 T^2}{128\pi^2 m_\chi}=\frac{g^3 T^2}{128\pi^2
 |\varphi|}}\lesssim \frac{g^3 T}{128\pi^2},
\qquad~ & \mbox{ for~} |\varphi| \gtrsim T,  \qquad
\end{array}
\right. \label{183} \\[5mm]
 \Gamma_M &\cong& \lnk\begin{array}{ll}
\displaystyle{\frac{g^4\varphi^2}{128\pi M} =
 \frac{g^3\varphi^2}{128\pi T} 
\lesssim \frac{g^3 T}{128\pi}}, \qquad\qquad & \mbox{ for~} |\varphi| <
T, \\[5mm] 
\displaystyle{\frac{f^4\varphi^2}{128\pi M} =  \frac{gf^2 T}{128\pi}}
\lesssim \frac{f^4\varphi^2}{128\pi gT}, 
 \qquad\qquad & \mbox{ for~} fT < 
f|\varphi| < gT.
 \label{gM}\end{array}
\right. 
\eea
In the second equality of (\ref{183}) we have put $m_\chi=g|\varphi|$.
If two or more channels are at work, the total dissipation rate is given
by their sum.

Next for $f|\varphi| < T < g|\varphi| ~(< |\varphi|)$, 
fields coupled to $\varphi$ with gauge coupling
strength are not thermalized and only those coupled with Yukawa coupling
strength are relevant.  Hence when $H < fT$, the scalar field oscillates
with the angular frequency $M=fT \ll T$.  In this case only the
following
two channels could be nonvanishing.
\bea
 \Gamma_B &\cong& \frac{(f\mu)^2 T}{2\pi M^2}=\frac{\mu^2}{2\pi T},
\qquad\qquad\qquad {\rm for~} \mu |\varphi| < fT^2, \label{185} \\[5mm]
 \Gamma_S &\cong& \frac{f^3 T^2}{128\pi^2 |\varphi|}
\lesssim \frac{gf^3T}{128\pi^2}. \label{186}
\eea

\subsection{Dissipation rate at the onset of field oscillation}

Finally we combine the above results with the thermal history and
the initial condition of $\varphi$ after inflation in order to evaluate
the dissipation rate at the onset of field oscillation.
After inflation, $\varphi$ is expected to trace the instantaneous minimum
(\ref{varphiexpect}) $|\varphi| \simeq (H\mst^{n-3})^{1/(n-2)}$ until
the onset of field oscillation due to a thermal mass. 
For definiteness let us consider the case preheating is not effective
so that  the cosmic temperature is given by
(\ref{temperature2}), $T \simeq (T_R^2H\mg)^{1/4}$, 
during the inflaton field oscillation regime.  Let us also take $n=4$
below.   

The flat direction starts oscillation with a frequency $gT$ if both 
$T> g|\varphi| $ and $gT > H$ hold true, or with a frequency
$fT$ when both $T > f|\varphi|$ and $fT>H$ hold, whichever comes
earlier.  The condition $T > g|\varphi|$ is satisfied when
\beq
  H < g^{-4}T_R^2\mst^{-2}\mg \equiv H_{\rm th,g},
\eeq
while $T > f|\varphi|$  applies when
\beq
  H < f^{-4}T_R^2\mst^{-2}\mg \equiv H_{\rm th,f}.
\eeq
On the other hand, the inequality $gT >H$ holds when
\beq
 H < g^{4/3}T_R^{2/3}\mg^{1/3}\equiv  H_{\rm s,g},
\eeq
while $fT>H$ is fulfilled when
\beq
 H < f^{4/3}T_R^{2/3}\mg^{1/3}\equiv  H_{\rm s,f}.
\eeq
We find $H_{\rm s,f}<H_{\rm s,g}$ and $H_{\rm th,g}<H_{\rm th,f}$.

The flat direction starts oscillation with the frequency $M=gT$ at
$H=\min(H_{\rm s,g},H_{\rm th,g}),$ if 
$\min(H_{\rm s,g},H_{\rm th,g})>\min(H_{\rm s,f},H_{\rm th,f})$.
This inequality holds true if
\beq
 T_R > g^3f \mst^{3/2}\mg^{-1/2}=8\times 10^{10}\lmk\frac{g}{0.5}\rmk^3
\lmk\frac{f}{10^{-3}}\rmk\lmk\frac{\mst}{10^{16}\GeV}\rmk^{3/2}\GeV.
\eeq
In this case the dissipation rate is given by (\ref{gF}) through
(\ref{gM}) depending on the value of $|\varphi|$ and $T$ then.

On the other hand, if 
$\min(H_{\rm s,g},H_{\rm th,g})<\min(H_{\rm s,f},H_{\rm th,f})$,
the scalar field starts oscillation with the frequency $M=fT$
at $H=\min(H_{\rm s,f},H_{\rm th,f}).$  This happens if
\beq
 T_R < g^3f \mst^{3/2}\mg^{-1/2},
\eeq
and the dissipation rate is given by (\ref{185}) or (\ref{186}).

To conclude we have calculated the evaporation rate of the flat
direction at the onset of its oscillation for $n=4$.  We find
that in some cases the rate may be larger than the previous estimate
based on particle-particle scattering picture (\ref{scat}) but the time
scale of evaporation is long enough that significant  oscillation is
certainly possible before evaporation.  Once it starts oscillation, the
evolution of the field amplitude $|\varphi|$ becomes different from 
(\ref{varphiexpect}), so one must solve its evolution together with the
thermal history after inflation in order to determine when the flat
direction completes thermalization.  This issue will be analyzed
elsewhere together with the amount of baryon asymmetry produced, 
where two-loop logarithmic correction to the effective
potential \cite{Anisimov:2000wx}, 
which we have neglected here, will also be included.  

\section{Discussion}

In the present paper we have developed a formalism to investigate the
relaxation processes of an oscillating scalar field $\phi$ interacting
with various particles in a thermal state using the in-in formalism of
nonequilibrium quantum field theory.  Integrating out those thermal
particles interacting with $\phi$, we have obtained the effective action for
$\phi$ which is complex even if it is a real scalar field.  This is a
result of coarse-graining and manifestation of the dissipative effect on
$\phi$ to those integrated out.  The real equation of motion is
obtained by introducing auxiliary fields, $\xi_a(x)$ and $\xi_m(x)$,
 which act as an additive and a multiplicative noise
term, respectively.  
The former originates from interactions 
linear in $\phi$ such as Yukawa coupling or three-body bosonic
interaction, while the latter is from quadratic or higher-order
interactions in $\phi$. It induces noises on the
effective mass of $\phi$.

The equation of motion has terms nonlocal in both space and
time as a result of quantum corrections.  In the linear regime when
higher-order terms in $\phi$ are negligible in the equation of motion,
these nonlocalities can easily be handled because its
 Fourier modes are decoupled from each other.  As a result we
can find an analytic solution for each mode from which we
can extract the dissipation rate.  On the other hand, the dissipation
rates from multiplicative interactions are read  from the equation of
motion itself.

Quite generally, the memory kernels, which generate nonlocal terms in the
equation of motion, are determined by the imaginary part of the Green
functions relevant to each diagram, while 
the noise correlation functions are identical to the real part of the
same function up to a numerical factor.  We have found that for all the
interactions discussed here, the Fourier transform of the memory kernel
and that of noise correlation function take a specific ratio which is
determined only by the temperature and the angular frequency of the
mode.  This relation is achieved by microphysical detailed balance
relation.  It also 
leads to the well-known
fluctuation-dissipation theorem  for low-momentum modes,
which guarantees that the scalar field
relaxes to a state the equipartition law is satisfied.
For higher-momentum modes the scalar field  relaxes to the  thermal
equilibrium state with the same temperature
 where the number density of each quanta consists of
the boson distribution function and its zero-point vacuum component.

Although we have shown 
the fate of the oscillating scalar field is the same
equilibrium state, the time scale of relaxation to it is strikingly
different depending on the nature of interactions.  In the case of
Yukawa coupling with fermions, the dissipation rate takes a smaller value
at finite temperature than the zero-temperature decay width due to the
Pauli blocking.  On the other hand, in the case of bosonic three-body
interaction, the dissipation rate is larger than the zero temperature
decay rate due to the induced effect.  
As a result we have seen the reheat temperature after inflation may be
somewhat changed from conventional estimates, and that 
in an extreme case the inflaton can dissipate its energy even 
 without linear interactions that leads to its decay.  

The temperature
dependence on the dissipation rate may also affect on the property and
the spectrum of density fluctuations.  It has been known for a long time
that primordially isocurvature fluctuations that were stored in a long-lived
scalar field during inflation when it was subdominant are converted to
the adiabatic ones as its energy density tends to dominate the Universe
later 
\cite{Mollerach:hu}.  Such a property has been utilized in some models
of non-scale invariant fluctuations
\cite{Kofman:1986wm,Yokoyama:1995ex}. 
Nowadays the above
conversion mechanism from isocurvature to adiabatic fluctuations is
called the curvaton scenario  \cite{Lyth:2002my}.
When the curvaton field decays, the dependence of their dissipation rate
on the background temperature may induce additional fluctuations just as
in the modulated coupling scenario \cite{Dvali:2003em}.
On the other hand, a model of baryogenesis has been proposed in which
small fluctuation in the inflaton's dissipation rate induces enhanced
baryon-number fluctuations \cite{Yokoyama:1987he}.  
These possible effects on density fluctuations due to the temperature
dependence on the dissipation rate will be studied elsewhere.

The dissipation associated
with interactions linear in $\phi$ such as Yukawa coupling and
three-body bosonic coupling can be interpreted in terms of  decay, while
the setting-sun diagram from biquadratic coupling $h^2\phi^2\chi^2/4$
and quartic coupling $\lambda\phi^4/4!$
induces dissipation associated with scattering.  The former is 
suppressed when the would-be decay products are more massive than the
oscillation frequency, but the latter is effective even in this
regime.  Consequently the dissipation rate of the Affleck-Dine flat
direction field shows a rather complicated behavior depending on the
evolution of its oscillation amplitude and the temperature.  Although the
dissipation time scale is much longer than the oscillation period,
whether sufficient baryon number is generated or not depends on the
magnitude of the A terms as well, in particular on the presence or the
absence of the thermal A term \cite{Anisimov:2000wx}.
  Hence the both ingredients should be
analyzed properly to yield final baryon asymmetry.

In the present paper we have  concentrated on the fate of the
zero-mode oscillation but we can easily obtain the dissipation rates of
higher-momentum modes using our formula and we expect they have larger
dissipation rate.  This may affect formation of $Q$-balls.

Thus there are a number of interesting problems associated with
dissipation of flat directions remaining. 
 We hope to return to these issues in
near future.

\acknowledgments{The author is grateful to Sasha Dolgov and Rocky Kolb
for discussions.  
This work was partially supported by the JSPS
  Grant-in-Aid for Scientific Research No.\ 16340076.}

\end{document}